\documentclass[12pt,a4paper]{iopart}
\usepackage{iopams}
\usepackage[english]{babel}
\usepackage{float}
\usepackage[dvips]{graphicx}
\usepackage{rotating}
\usepackage{epsfig}
\usepackage[dvips]{color}
\definecolor{Red}{named}{Red}
\def\cnub{C$\nu$B}
\newcommand{\lwig}{\mbox{\;\raisebox{.3ex}
    {$<$}$\!\!\!\!\!$\raisebox{-.9ex}{$\sim$}\;}}
\newcommand{\gwig}{\mbox{\;\raisebox{.3ex}
    {$>$}$\!\!\!\!\!$\raisebox{-.9ex}{$\sim$}}\;}

\begin{document}
\title{The Cosmic Neutrino Background Anisotropy - Linear Theory}
\author{Steen Hannestad$^1$, Jacob Brandbyge$^1$}
\ead{sth@phys.au.dk, jacobb@phys.au.dk}
\address{$^1$Department of Physics and Astronomy, University
of Aarhus, Ny Munkegade, DK-8000 Aarhus C, Denmark}
\date{\today}

\begin{abstract}
The Cosmic Neutrino Background ({\cnub}) anisotropy is calculated for massive neutrino states by solving the
full Boltzmann equation. The effect of weak gravitational lensing, including the Limber approximation, is also derived for massive particles, and
subsequently applied to the case of massive neutrinos.
\end{abstract}

\section{Introduction}

The Cosmic photon Microwave Background (CMB) is currently our main source of information about the physical
content of the Universe. Observations of the CMB anisotropy provides detailed information about the curvature
of the Universe, the matter content, and a plethora of other parameters \cite{Komatsu:2008hk}.

Standard model
physics likewise predicts the presence of a Cosmic Neutrino Background (\cnub) with a well defined
temperature of $T_\nu \sim (4/11)^{1/3} T_\gamma$. While it remains undetected in direct experiments, the
presence of the {\cnub} is strongly hinted at in CMB data. The homogeneous C$\nu$B component has been
detected at the 4-5$\sigma$ level in the WMAP data (see e.g.\ \cite{Komatsu:2008hk,Hamann:2007pi,de
Bernardis:2007bu,Ichikawa:2008pz,Hamann:2008we,Popa:2008tb}). Furthermore, this component is known to be
free-streaming, i.e. to have an anisotropic stress component consistent with what is expected from standard
model neutrinos (see
\cite{Bashinsky:2003tk,Trotta:2004ty,Bell:2005dr,DeBernardis:2008ys,Basboll:2008fx,Hannestad:2004qu,Friedland:2007vv}).
Finally the standard model neutrino decoupling history is also confirmed by Big Bang Nucleosynthesis (BBN),
the outcome of which depends on both the energy density and flavour composition of the {\cnub}.

While this indirect evidence for the presence of a {\cnub} is important, a direct detection remains an
intriguing, but almost impossible goal. The most credible proposed method is to look for a peak in beta
decay spectra related to neutrino absorption from the {\cnub} \cite{Weinberg:1962zz,Cocco:2007za,Blennow:2008fh}, although
many other possibilities have been discussed
\cite{Weiler:1982qy,Stodolsky:1974aq,Gelmini:2004hg,Ringwald:2004np,Fodor:2002hy,Duda:2001hd,Langacker:1982ih,Cabibbo:1982bb}.
The neutrino absorption method was first investigated by Weinberg \cite{Weinberg:1962zz}, based on the
possibility that the primordial neutrino density could be orders of magnitude higher than normally assumed
due to the presence of a large chemical potential. Although a large chemical potential has been ruled out
because it is in conflict with BBN and CMB
\cite{Pastor:2001iu,Pastor:2008ti,Simha:2008mt,Wong:2002fa,Abazajian:2002qx}, the method may still work and
recently there has been renewed interest in detecting the {\cnub} using beta unstable nuclei.

Although the direct detection of the {\cnub} is already very challenging, one might speculate on the
possibility that in the more distant future anisotropies in the {\cnub} will be detectable. For massless
neutrinos the calculation of the {\cnub} proceeds in a way which is almost identical to the standard CMB
calculation. The massless {\cnub} anisotropy spectrum was first presented in \cite{Hu:1995fqa} and
subsequently calculated in \cite{Michney:2006mk} in a highly simplified way which contains some, but not all
of the essential physics.

For massive neutrinos the calculation is much more complicated, and the {\cnub} anisotropy is changed
considerably: If the mass is sufficiently high, neutrino velocities can be as low as the escape velocities of
galaxies. In this case the {\cnub} is entirely determined by non-linear gravitational clustering. The current
thermal velocity of a non-relativistic homogeneous neutrino background is given roughly by $\langle v \rangle
\sim 1500 \, {\rm km} \, {\rm s}^{-1} \left(\frac{0.1 \, {\rm eV}}{m_\nu}\right)$, which should be compared
to gravitational streaming velocities which are up to $\sim 1000 \, {\rm km} \, {\rm s}^{-1}$. For small
masses (i.e.\ non-degenerate, $m_\nu \lwig 0.1$ eV) it is possible to make a calculation which is analogous
to what is done for the CMB. As will be explained later the {\cnub} spectrum is shifted to larger angular
scales, mainly because the much shorter distance to the neutrino last scattering surface changes the relation
between angular scales and length scales. Furthermore the amplitude of the anisotropy greatly increases at
small multipoles because the gravitational source term becomes much more important as neutrinos become
increasingly non-relativistic at late times.

In addition to this change in the primary {\cnub} spectrum, the effect of gravitational lensing is also very
different from the case of massless neutrinos. As is the case for the primary spectrum, gravitational lensing
becomes increasingly important at low $l$ as the mass increases. A detailed calculation of the lensing of
massive neutrinos is presented in section 3. First, however, we derive the necessary equations for the
primary {\cnub} anisotropy in section 2 and present a numerical calculation of the {\cnub} power spectra for
different masses. In section 4 we combine the results of sections 2 and 3 to derive the form of the lensed
massive {\cnub}. Finally, section 5 contains a discussion of our results and our conclusions.

\section{The primary C$\nu$B}

\subsection{Theory - The Boltzmann equation}

The evolution of any given particle species can be described via the Boltzmann equation. Our notation is
similar to that of Ma and Bertschinger \cite{MB}. As the time variable we use conformal time, defined as $d
\eta \equiv dt/a(t)$, where $a(t)$ is the scale factor and $t$ is cosmic time. Also, as the momentum variable we shall use the
comoving momentum $q_j \equiv a p_j$. We further parametrise $q_j$ as $q_j = q n_j$, where $q$ is the
magnitude of the comoving momentum and $n_j$ is a unit 3-vector specifying direction.

The Boltzmann equation can generically be written as
\begin{equation}
L[f] = \frac{Df}{D\tau} = C[f],
\end{equation}
where $L[f]$ is the Liouville operator. The collision operator $C[f]$ on the right-hand side describes any
possible collisional interactions. For neutrinos $C[f]=0$ after neutrino decoupling at $T \sim 2-3$ MeV.

We then write the distribution function as
\begin{equation}
f(x^i,q,n_j,\eta) = f_0(q) [1+\Psi(x^i,q,n_j,\eta)], \label{eq:Psi}
\end{equation}
where $f_0(q)$ is the unperturbed distribution function. For a fermion which decouples while relativistic,
this distribution function is
\begin{equation}
f_0(q) = [\exp(q/T_0)+1]^{-1},
\end{equation}
where $T_0$ is the present-day temperature of the species.

In conformal Newtonian (longitudinal) gauge the Boltzmann equation for neutrinos can be written as an evolution equation
for $\Psi$ in $k$-space \cite{MB}
\begin{equation}
\frac{1}{f_0} L[f] = \frac{\partial \Psi}{\partial \tau} + ik \frac{q}{\epsilon} \mu \Psi + \frac{d \ln
f_0}{d \ln q} \left[\dot{\phi}-ik \frac{\epsilon}{q} \mu \psi \right] = 0,
\label{eq:boltzmann}
\end{equation}
where $\mu \equiv n^j \hat{k}_j$. $\psi$ and $\phi$ are the metric perturbations, defined from the perturbed
space-time metric in the conformal Newtonian gauge \cite{MB}
\begin{equation}
ds^2 = a^2(\eta) [-(1+2\psi)d\eta^2 + (1-2\phi)\delta_{ij} dx^i dx^j].
\end{equation}

The perturbation to the distribution function can be expanded as follows
\begin{equation}
\Psi = \sum_{l=0}^{\infty}(-i)^l (2l+1)\Psi_l P_l(\mu).
\end{equation}
One can then write the collisionless Boltzmann equation as a moment hierarchy for the $\Psi_l$'s by
performing the angular integration of $L[f]$
\begin{eqnarray}
\dot\Psi_0 & = & -k \frac{q}{\epsilon} \Psi_1 + \dot{\phi} \frac{d \ln f_0} {d \ln q}, \label{eq:psi0}\\
\dot\Psi_1 & = & k \frac{q}{3 \epsilon}(\Psi_0 - 2 \Psi_2) - \frac{\epsilon k}{3 q} \psi \frac{d \ln f_0}{d
\ln q}, \label{eq:psi1}\\ \dot\Psi_l & = & k \frac{q}{(2l+1)\epsilon}(l \Psi_{l-1} - (l+1)\Psi_{l+1}) \,\,\,
, \,\,\, l \geq 2.
\end{eqnarray}
By integrating the neutrino perturbation over momentum
\begin{equation}
F_{\nu l} = \frac{\int dq q^2 \epsilon f_0(q) \Psi_l}{\int dq q^2 \epsilon f_0(q)},
\end{equation}
one finds a set of equations equivalent to those used to follow the perturbations in photons or massless
neutrinos, i.e\ a set of equations for $F_{\nu l}$.

The distortion to the sky intensity can also be found using the perturbation to the temperature, $\Theta$,
related to the distribution function via
\begin{equation}
f(q) = [\exp(q/[T_0(1+\Theta)]+1]^{-1}. \label{eq:Delta}
\end{equation}
Equating Eqs.~(\ref{eq:Psi}) and (\ref{eq:Delta}), $\Psi$ and $\Theta$ are related by
\begin{equation}
\Theta(q) = -\left(\frac{d {\rm ln} f_0}{d {\rm ln} q}\right)^{-1} \Psi(q).
\end{equation}
Since the transformation between $\Psi$ and $\Theta$ is mass-independent, the normalisation of the angular
power spectrum is not significantly affected. \cite{Michney:2006mk} used the transformation
$-\left(\frac{d {\rm ln} f_0}{d {\rm ln} \epsilon}\right)^{-1}$ which introduces an extra factor
of $v^2$ relative to our definition. The $v^2$ factor, by construction, suppresses the perturbations
for the higher neutrino masses significantly.

By substituting $\Theta$ into the Boltzmann equation, Eq.~(\ref{eq:boltzmann}), it can be seen that $\Theta$
is $q$-independent in the massless case. For massless particles it is therefore convenient to calculate the
angular power spectrum of the temperature perturbation. For massive particles $\Theta$ is $q$-dependent
giving rise to spectral distortions in the temperature field. Since it is convenient to have results similar
to the CMB in the massless neutrino case we calculate the massive neutrino anisotropy spectrum of the
quantity $\Theta_l (q)$ related to $\Theta_{\nu l}$ by
\begin{equation}
\Theta_{\nu l} = \frac{\int dq q^2 \epsilon f_0(q) \Theta_{l}(q)}{\int dq q^2 \epsilon f_0(q)}.
\end{equation}

Of course, the quantity which is actually measurable will depend on the type of experiment used. A typical
experiment will measure either a number flux, a momentum flux, or a kinetic energy flux as a function of
angle. For massless particles these quantities are trivially related by a momentum independent number, i.e.\
the number flux anisotropy is $3 \Theta_{\nu l}$ and the momentum/energy flux anisotropy is $F_{\nu l} = 4
\Theta_{\nu l}$. For massive particles this is no longer true and one must calculate the appropriate quantity
for any given experiment.

Similarly to the photon case one can then construct the {\cnub} sky brightness fluctuation angular power
spectrum as
\begin{equation}
C_l^\Theta(q) = (4 \pi)^2 \int k^2 dk P_I(k) \Theta_l^2(q,k).
\end{equation}
Here $P_I(k)$ is the primordial potential fluctuation power spectrum, $P_I(k) \propto k^{n-4}$. Throughout
the paper we assume a flat Harrison-Zel'dovich spectrum so that $n=1$ with cosmological parameters
$(\Omega_{\rm b},~\Omega_{\rm m},~\Omega_{\rm \Lambda},~h,~A_{\rm
s})~=~(0.05,~0.3,~0.7,~0.7,~2.3\cdot10^{-9})$. As can be seen, a given $l$ gets contributions from all $k$.
The extra $q$-dependence arises because in principle one should perform the $q$-dependent lensing of
$C_l^\Theta(q)$ in $q$-bins. The total $C_l^\Theta$ is found by averaging over momenta at the present time
$\eta_0$
\begin{equation}
C_l^\Theta = \left[\frac{\int dq q^2 \epsilon(\eta_0,q) f_0(q) \sqrt{C_l^\Theta(q)}}{\int dq q^2
\epsilon(\eta_0,q) f_0(q)}\right]^2.
\end{equation}
$C_l^\Theta$ does not include lensing, since the second-order term encoding deflections has been left out of
the Boltzmann equation. Note that mapping the neutrino anisotropic sky in different momentum bins
(anisotropic neutrino momentum tomography) will probe structures at different spacial distances from us.
Furthermore, the observed C$\nu$B will be a superposition of the spectra for each individual neutrino mass.

\begin{figure}
   \noindent
      \begin{center}
      \hspace*{-0.1cm}\includegraphics[width=0.8\linewidth]{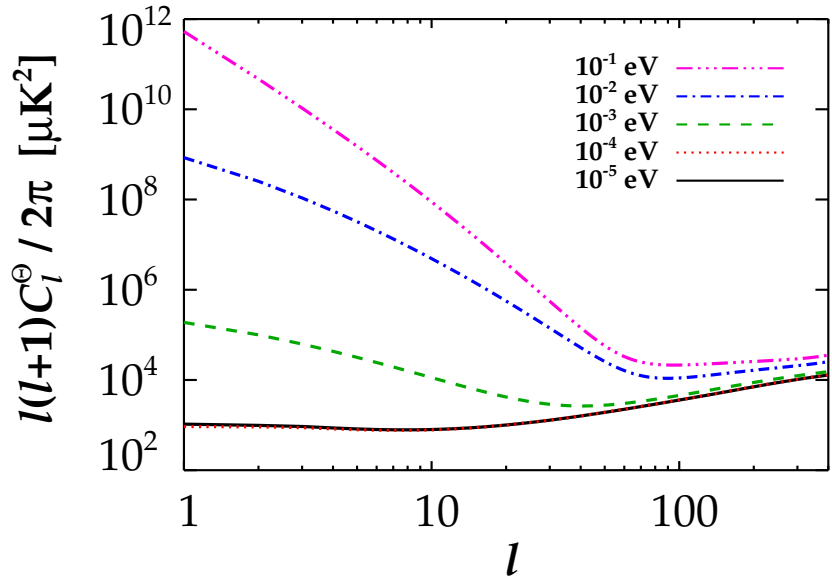}
      \end{center}
   \caption{Primary {\cnub} spectrum for different neutrino masses.}
   \label{fig:transfer}
\end{figure}

\subsection{Gauge effects}
As noted we use the conformal Newtonian gauge, since this gauge is directly related to
physically measurable quantities. In the Synchronous gauge the velocity perturbation, $\theta$, for the CDM
component is, by definition, zero. Therefore $\theta_\nu$, which is a momentum integral over $\Psi_1$, is a
gauge dependent quantity. In contrast, the anisotropic stress, which is a momentum integral over $\Psi_2$,
is a gauge independent quantity. Since all moments $\Psi_l$ with $l>2$ is recursively related to $\Psi_2$,
these higher order moments are gauge independent as well.

Since we cannot separate the CMB/{\cnub} dipole from our own peculiar motion, we are only interested in
modelling the $C_l^\Theta$'s with $l \geq 2$ when comparing with observations. But $C_1$, the lowest mode
containing physically relevant information, is gauge dependent. We have taken this into account by working
in the physical conformal Newtonian gauge.

We also note that the transfer functions are gauge dependent, though for the Synchronous and conformal
Newtonian gauges they are almost identical inside the horizon for the massive components. Therefore we have
calculated the transfer functions used to get the lensing contribution in the Synchronous gauge with CAMB
\cite{CAMB}.

\begin{figure}
   \noindent
  \begin{minipage}{1.0\linewidth}

     \begin{minipage}{0.49\linewidth}
        \begin{turn}{90}
        \hspace*{-0.0cm}\includegraphics[width=0.62\linewidth]{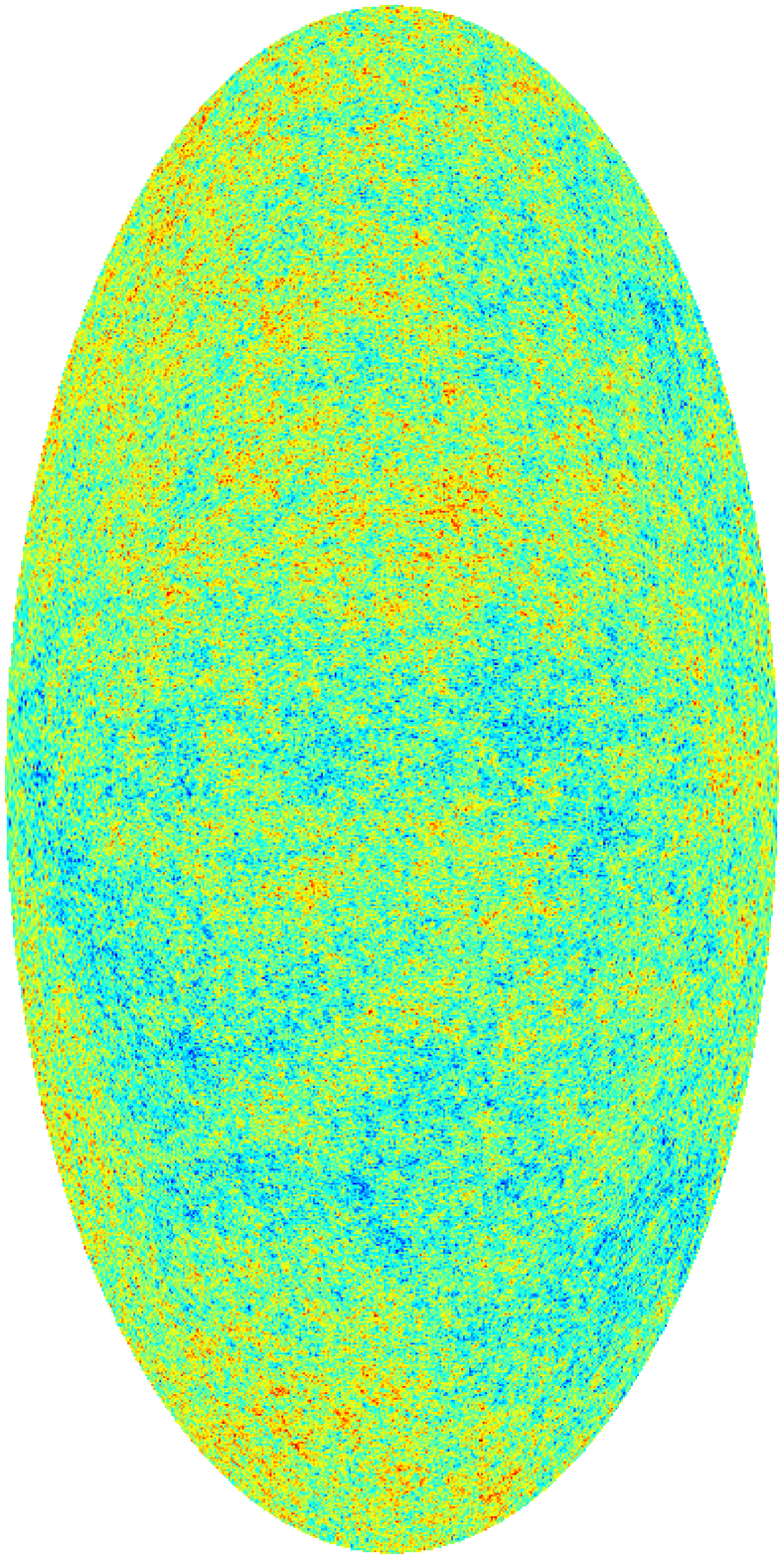}
        \end{turn}
     \end{minipage}
     \begin{minipage}{0.49\linewidth}
        \begin{turn}{90}
       \hspace*{-0cm}\includegraphics[width=0.62\linewidth]{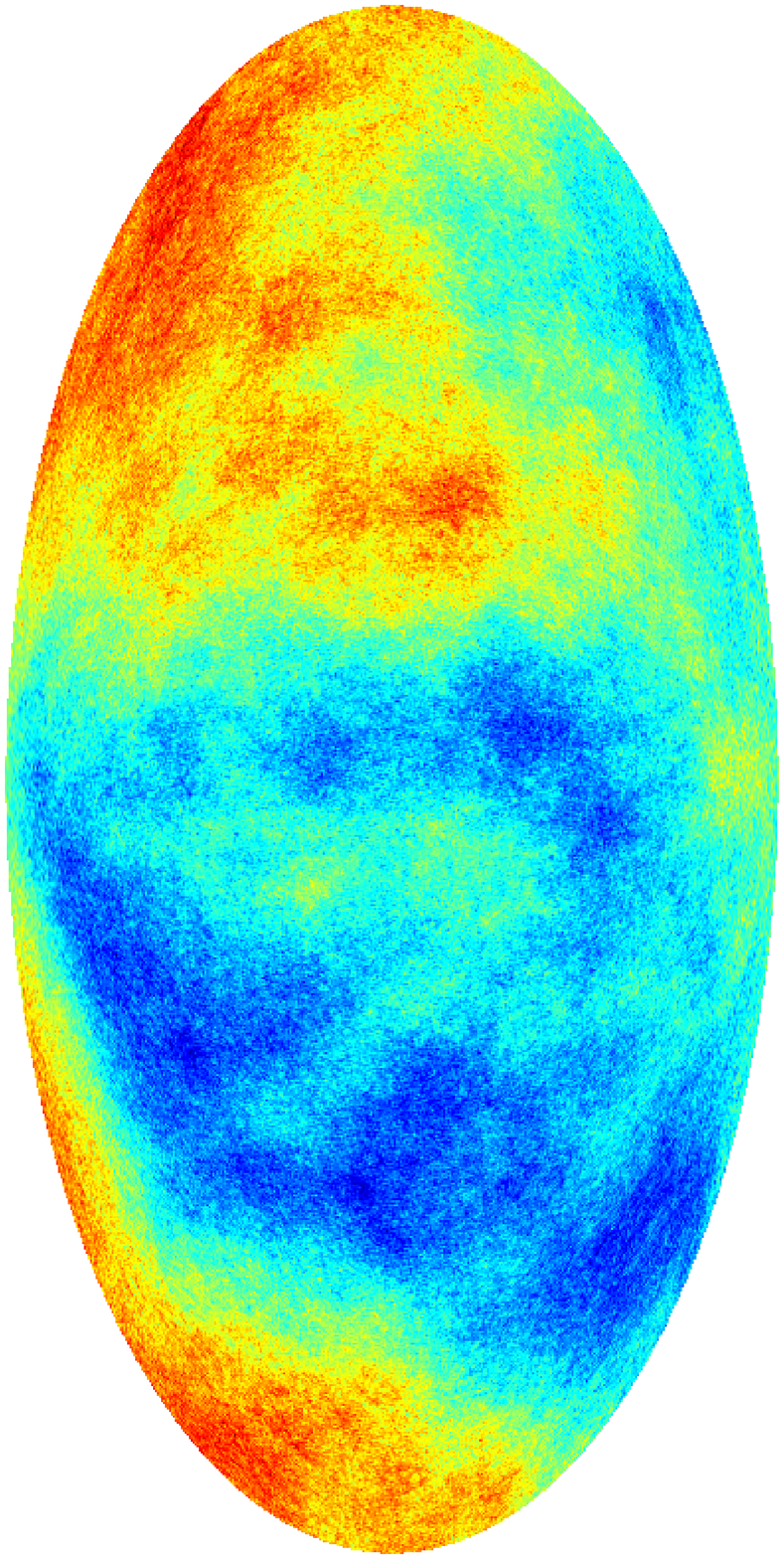}
          \end{turn}
     \end{minipage}
     \begin{minipage}{0.49\linewidth}
        \begin{turn}{90}
        \hspace*{-0.0cm}\includegraphics[width=0.62\linewidth]{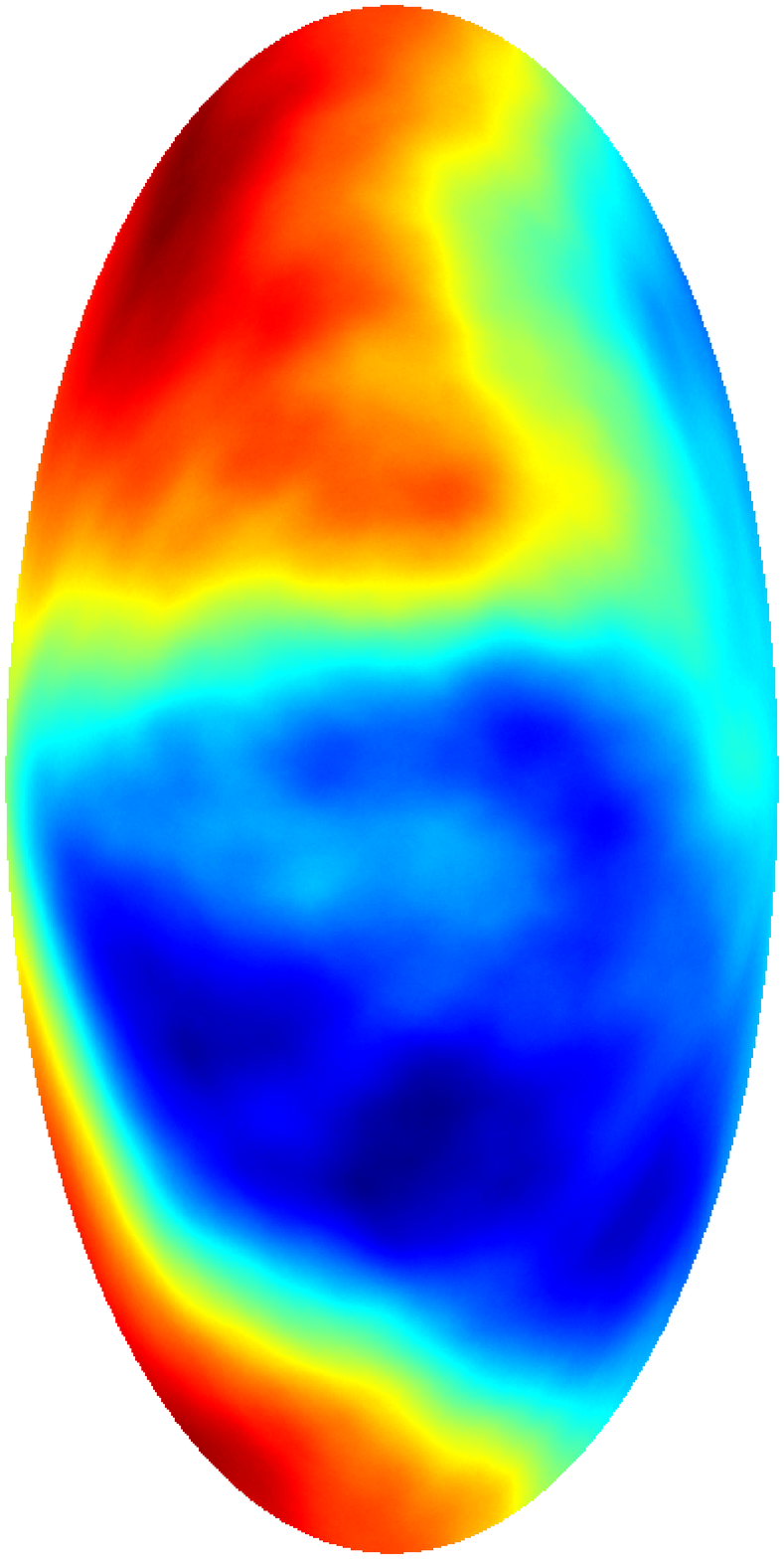}
        \end{turn}
     \end{minipage}
     \begin{minipage}{0.49\linewidth}
        \begin{turn}{90}
       \hspace*{-0cm}\includegraphics[width=0.62\linewidth]{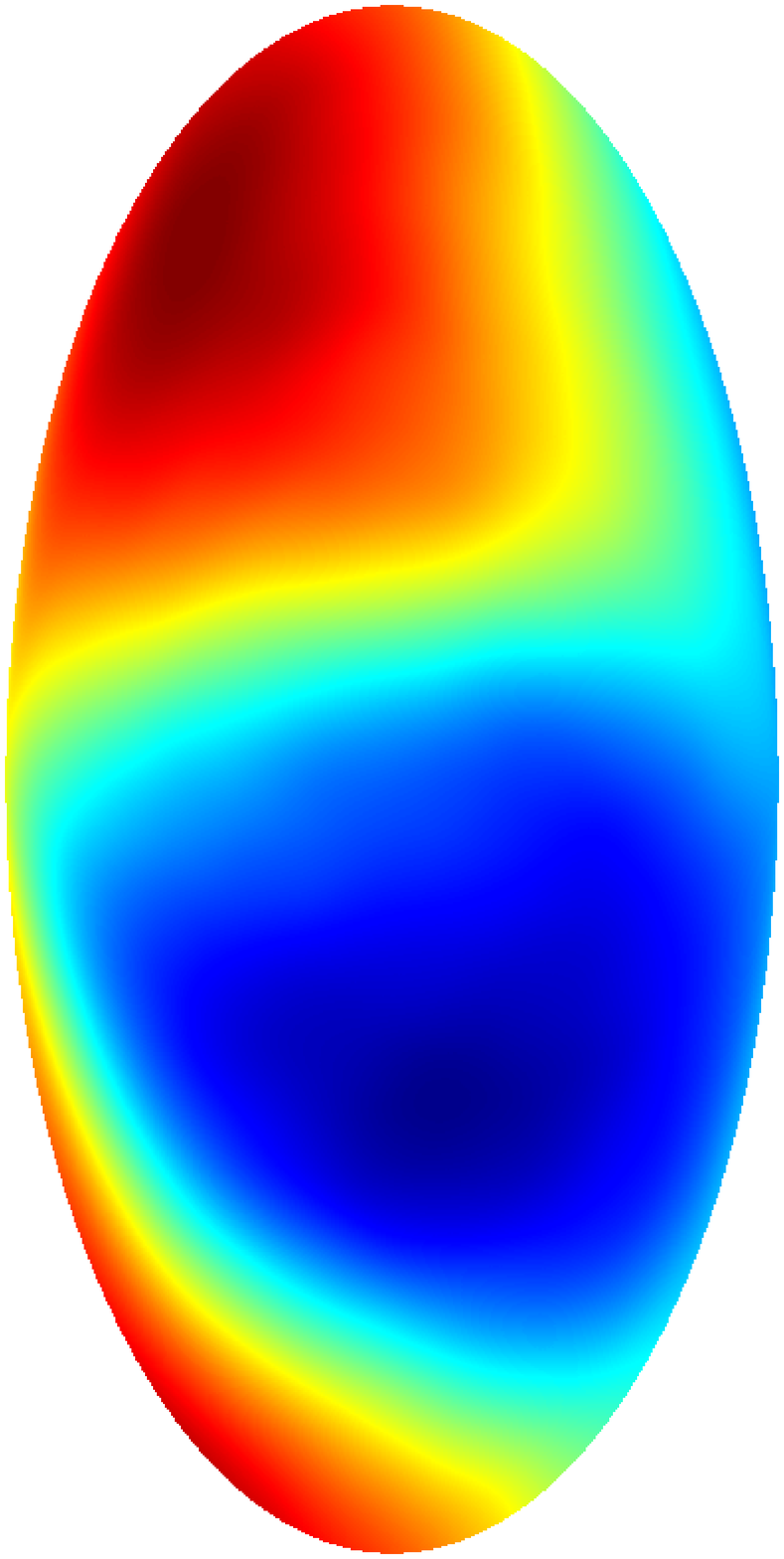}
          \end{turn}
     \end{minipage}

   \end{minipage}

   \caption{Sky maps of the primary neutrino power spectra, $C_l^\Theta$, with the dipole included, for $m_\nu = 10^{-5}$ eV
   (top-left), $10^{-3}$ eV (top-right), $10^{-2}$ eV (bottom-left) and $10^{-1}$ eV (bottom-right). The maps
   have been generated with the same underlying random numbers with the HEALPIX package
   \cite{Gorski:2004by}.}
   \label{fig:primary_sky}
\end{figure}

\begin{figure}
   \noindent
        \hspace*{-0.3cm}\includegraphics[width=1.1\linewidth]{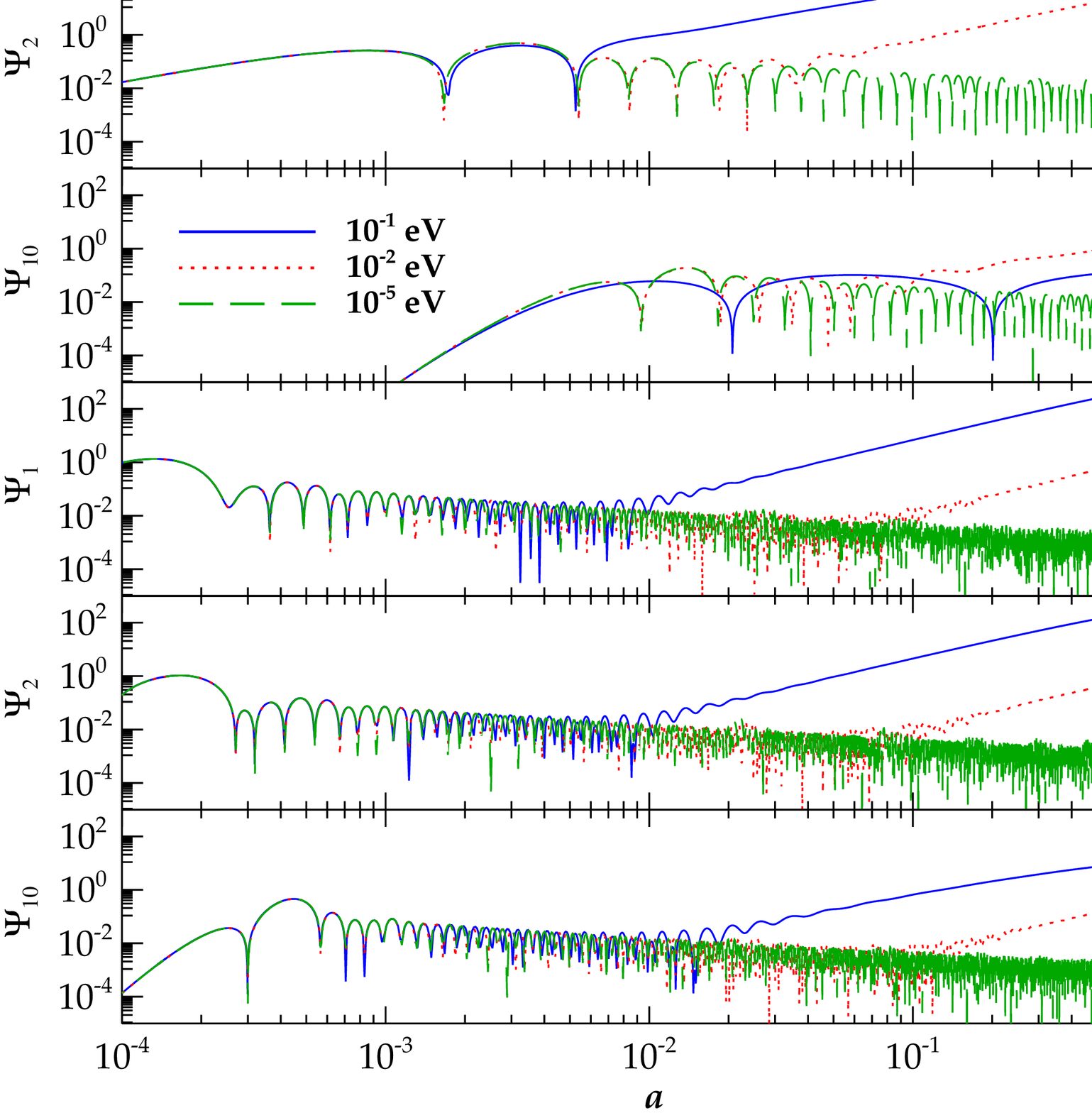}
        \vspace{1.0cm}
   \caption{$\Psi_l$'s for 3 neutrino masses with momentum $q/T_0=3$ as a function of the scale factor. The upper
   three panels are for $k = 0.01 \, h \, {\rm Mpc}^{-1}$ and the lower three panels for $k = 0.1 \, h \, {\rm
   Mpc}^{-1}$.}
   \label{fig:Psi_l}
\end{figure}

\subsection{Numerical results}

We have used the COSMICS code \cite{Bertschinger:1995er} to solve the Boltzmann hierarchy for the neutrinos. In
practise we have solved the system going up to $l=500$ with 64 bins in $q$, equally spaced from
$q/T_0~=$ 0 to 15. In Fig.~\ref{fig:transfer} we show results for $C_l^\Theta$ for various masses and Fig.~\ref{fig:primary_sky} shows sky map realisations for these spectra.

The massless case (i.e.\ $10^{-5}\, {\rm eV}$) is consistent with the result of \cite{Hu:1995fqa}. At high $l$ the spectra are
almost identical, and do not depend on the neutrino mass. The reason for this can be understood from the
following argument: Above a certain $k$-value, $k_{\rm FS}$, neutrinos are completely dominated by
free-streaming and this $k$-value is proportional to $m_\nu$. In order to convert this to an $l$-value one
then uses the relation $l_{\rm FS} \sim k_{\rm FS} \chi^*$ (where $\chi^*$ is the comoving coordinate from
which the neutrinos originate) and since $\chi^* \propto m_\nu^{-1}$ for non-relativistic particles \cite{Dodelson:2009ze},
$l_{\rm FS}$ does not depend on $m_\nu$. Inserting numbers one finds $l_{\rm FS} \sim 100$ which is in good
agreement with Fig.~\ref{fig:transfer}. At smaller angular scales, $l \gwig l_{\rm FS}$,
the anisotropy comes from the Sachs-Wolfe effect during radiation domination.

For smaller $l$-values the anisotropy increases dramatically as the mass increases. This can be understood as
follows. As soon as neutrinos go non-relativistic the $\frac{\epsilon k}{3 q} \psi \frac{d \ln f_0}{d \ln q}$
term in $\dot{\Psi}_1$ begins to dominate the Boltzmann hierarchy evolution. This quickly makes the higher
$l$ modes increase as well, and the final amplitude simply depends on the time elapsed after neutrinos go
non-relativistic.

The effect can be seen in Fig.~\ref{fig:Psi_l} which shows the evolution of $\Psi_1$,
$\Psi_2$ and $\Psi_{10}$ for three different neutrino masses and two different $k$-values. As soon as
neutrinos go non-relativistic $\Psi_1$ immediately begins to grow, and the higher $\Psi_l$'s follow with a
slight delay for $k = 0.1 \, h  \, {\rm Mpc}^{-1}$. This exactly matches the low $l$ behaviour seen in Fig.~\ref{fig:transfer}. Note also the
behaviour of $\Psi_{10}$ for $k=0.01 \, h \, {\rm Mpc}^{-1}$ and $m_\nu=0.1$ eV. For this mass and $k$-value, the
$q/\epsilon$ term in the $\dot{\Psi}_l$ equations becomes sufficiently important to suppress the propagation
of the gravitational source term to high $l$. At higher $k$ this is no longer true.

For high mass neutrinos ($\gtrsim 0.1 {\rm eV}$) gravitational distortion is so strong that the dominant
contribution comes from the local galactic and/or cluster halo. This case is similar to the study of the WIMP
flux anisotropy and requires $N$-body simulations \cite{Brandbyge1}. This will be treated separately in a
later paper while in the present paper we limit ourselves to the framework of linear theory.

\section{The lensing distortion}

\subsection{Theory}

For strictly massless neutrinos the lensing distortion is identical to that for photons \cite{Seljak:1995ve,Challinor1,Lewis1}. The change in angle
$d \alpha$ per unit path $d\chi$ is proportional to the transverse derivative of the gravitational potential,
$\psi$,
\begin{equation}
\frac{d \alpha}{d \chi} = -2 \nabla_\perp \psi,
\label{eq:massless}
\end{equation}
where $\nabla_\perp \psi$ is the component perpendicular to the line of sight. Relaxing the assumption of relativistic particles and
solving the geodesic equation for a neutrino propagating in a weak potential, one arrives at the result
\begin{equation}
\frac{d \alpha}{d \chi} = -\frac{1+v^2}{v} \nabla_\perp \psi.
\label{eq:deviation}
\end{equation}
Note that this result reduces to the ordinary Newtonian expression $\frac{d \alpha}{dx} = - \frac{1}{v} \nabla_\perp \psi$ in the limit $v \to 0$. It should also be
noted that the expression diverges as $v \to 0$ because the assumption $v > v_{\rm esc}$ is violated, i.e.\
particles with low velocity will be gravitationally bound in the potential.

From Eq.~(\ref{eq:deviation}) we can calculate the distortion spectrum in a manner similar to what is done
for the usual gravitational lensing spectrum. This is done with the assumption of Gaussian perturbations
and using the Born approximation \cite{Lewis1}.

The total deflection angle $\pmb{\alpha}$ is related to the lensing potential, $\Pi$, by
\begin{equation}
\pmb{\alpha}=\nabla_{\hat{{\bf n}}}\Pi.
\end{equation}
In a flat universe the normal formula for the angular lensing power spectrum, $\langle
\Pi_{lm}\Pi_{l'm'}^*\rangle=\delta_{ll'}\delta_{mm'}C_l^\Pi$, is given by
\begin{equation}
C_l^\Pi = 16 \pi \int \frac{dk}{k} \left[ \int_0^{\chi^*} d\chi \mathcal{P}^{1/2}_\psi(k,\eta_0-\chi) j_l(k
\chi) \left(\frac{\chi^*-\chi}{\chi^* \chi}\right)\right]^2,
\end{equation}
derived using Eq.~(\ref{eq:massless}). Here the power spectrum, $\mathcal{P}_\psi$, is related to the
ordinary matter power spectrum in the density contrast by $\mathcal{P}_\psi \propto a^{-2} P_m / k$. $\chi^*$
is the conformal distance at which the photons (or neutrinos) decoupled, taken to be a single source sphere,
and $j_l$ is a spherical Bessel function. $j_l(k \chi) (\chi^*-\chi)/(\chi^* \chi)$ is an effective window
function, which distributes power in $k$-space along the particle trajectory to angular $l$-space.

However, several changes are necessary when particles are allowed to have mass. The relation $d \chi = -d
\eta$ must be replaced with $d \chi = -v d \eta$ (the minus sign accounts for the fact that time and space
run in different directions, i.e. $\int_0^{\chi^*}{d\chi}\sim\int_{\eta_0}^{\eta^*}{d\eta}$ with
$\eta^*\simeq 0$, so that the observer is at the origin). In addition, the power spectrum
$\mathcal{P}_\psi(k,\eta_0-\chi)$ should be replaced by $\mathcal{P}_\psi(k,\eta)$. With these modifications,
the expression for massive particles becomes

\begin{equation}
C_l^\Pi(q) = 4 \pi \int \frac{dk}{k} [\Delta_l^\Pi(q,k)]^2,
\end{equation}
with
\begin{equation}
\Delta_l^\Pi(q,k) = \int_0^{\eta_0} d\eta\Delta_l^\Pi(q,k,\eta),
\end{equation}
and
\begin{equation}
\Delta_{l}^\Pi(q,k,\eta) = [1+v^2(q,\eta)] \mathcal{P}^{1/2}_\psi(k,\eta) j_l(k \chi(q,\eta))
\left[\frac{\chi^*(q)-\chi(q,\eta)}{\chi^*(q) \chi(q,\eta)}\right],
\end{equation}
where $v(q,\eta) = q/\epsilon = 1/\sqrt{1+a^2(\eta) m^2/q^2}$. $\chi^*$ is now momentum dependent since
neutrinos with different velocities have different distances to their respective last scattering surfaces,
though they still scattered at the same time $\eta^*$.

We calculate an average quantity of the lensing power spectrum found by doing an energy average over the
$C_l^\Pi (q)$'s
\begin{equation}
C_l^\Pi = \left[\frac{\int dq q^2 \epsilon(\eta_0,q) f_0(q) \sqrt{C_l^\Pi(q)}}{\int dq q^2 \epsilon(\eta_0,q)
f_0(q)}\right]^2.
\end{equation}

Using the orthogonality of the Bessel functions together with the fact that they pick out the scale $k\simeq
l/\chi$ at high $l$, the high $l$ limit of the above equations reduce to the Limber approximation (see e.g.\
\cite{kaiser,Jain:1996st}) for massive particles
\begin{equation}
C_l^\Pi(q) \simeq \frac{2\pi^2}{l^3} \int_0^{\eta_0}d\eta \frac{\chi}{v}[1+v^2]^2
\mathcal{P}_\psi(l/\chi,\eta) \left[\frac{\chi^*-\chi}{\chi^* \chi}\right]^2. \label{eq:limber}
\end{equation}
For $l\gtrsim 100$ this approximation is very good for all masses simulated.


\begin{figure}
   \noindent
      \begin{center}
      \hspace*{-0.1cm}\includegraphics[width=0.8\linewidth]{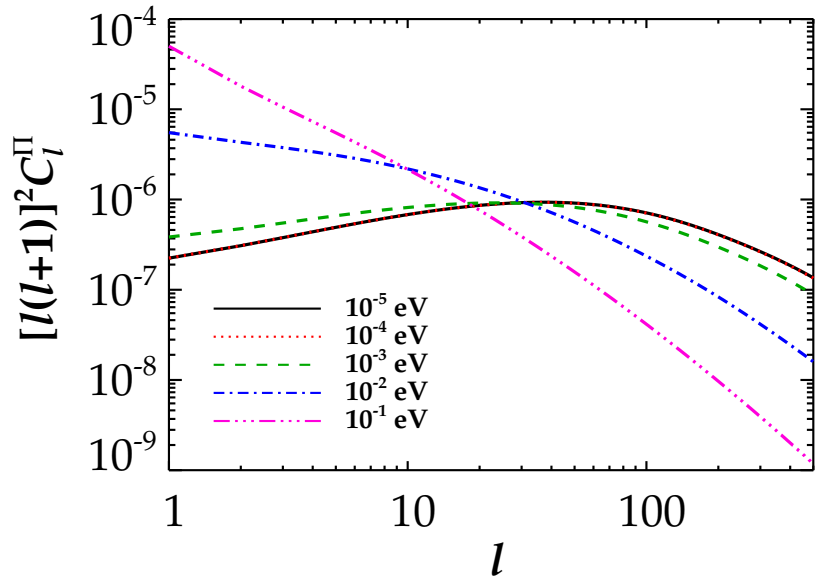}
      \end{center}
   \caption{The lensing potential angular power spectrum, $C_l^\Pi$, for 5 different neutrino masses.}
   \label{fig:lensing}
\end{figure}

\begin{figure}
   \noindent
  \begin{minipage}{1.0\linewidth}

     \begin{minipage}{0.49\linewidth}
        \begin{turn}{90}
        \hspace*{-0cm}\includegraphics[width=0.62\linewidth]{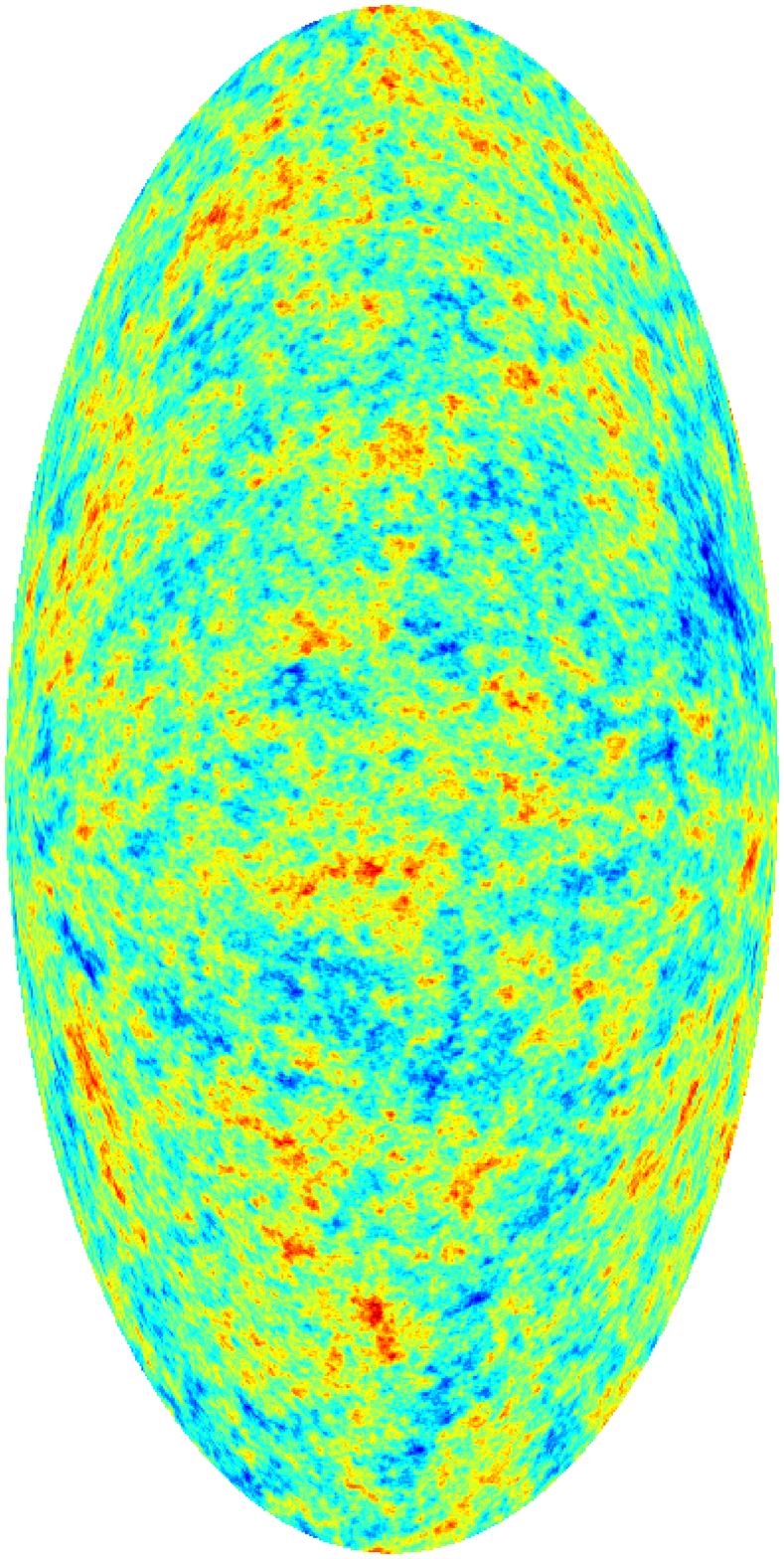}
        \end{turn}
     \end{minipage}
     \begin{minipage}{0.49\linewidth}
        \begin{turn}{90}
       \hspace*{-0cm}\includegraphics[width=0.62\linewidth]{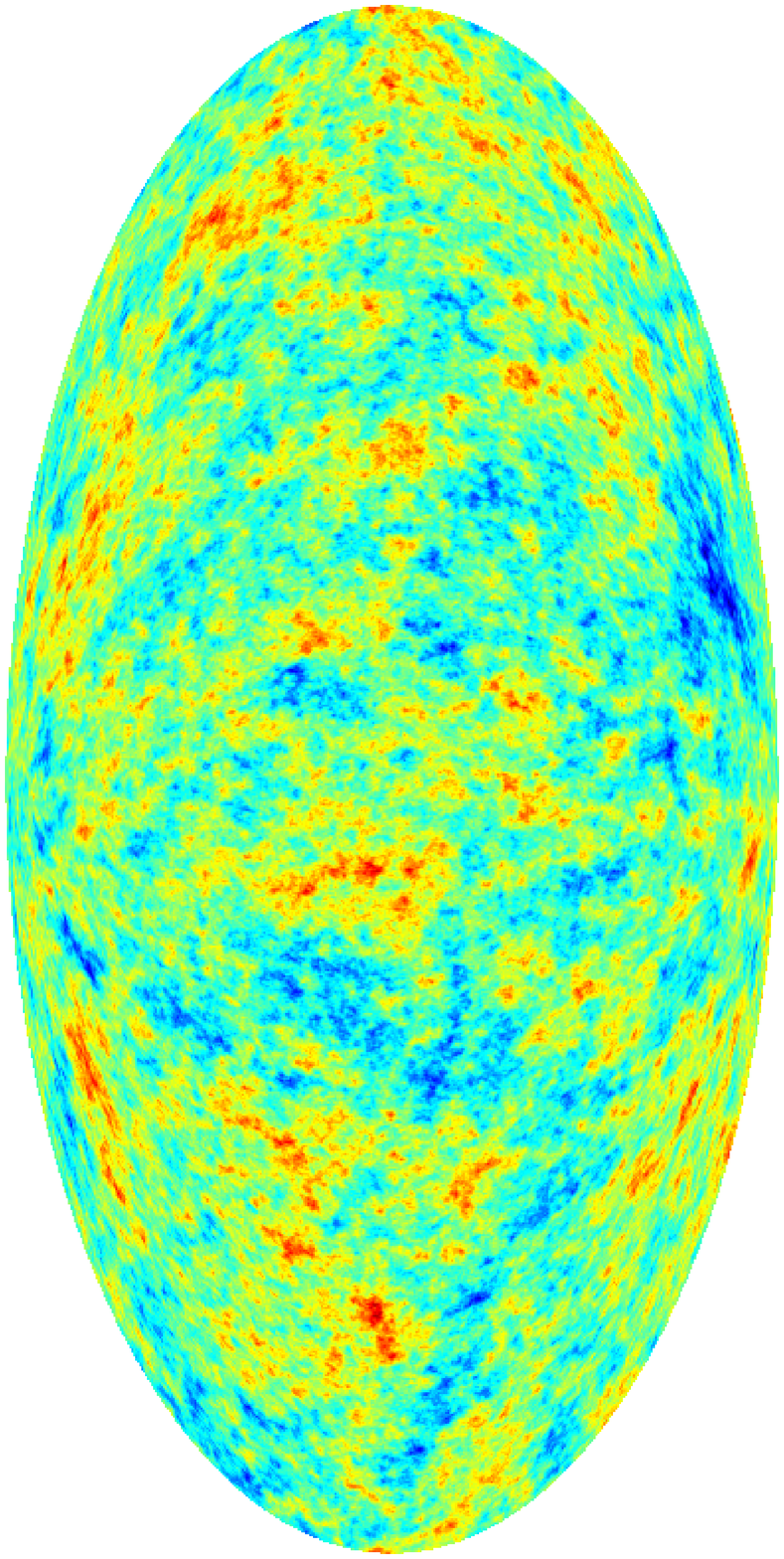}
          \end{turn}
     \end{minipage}
     \begin{minipage}{0.49\linewidth}
        \begin{turn}{90}
        \hspace*{-0cm}\includegraphics[width=0.62\linewidth]{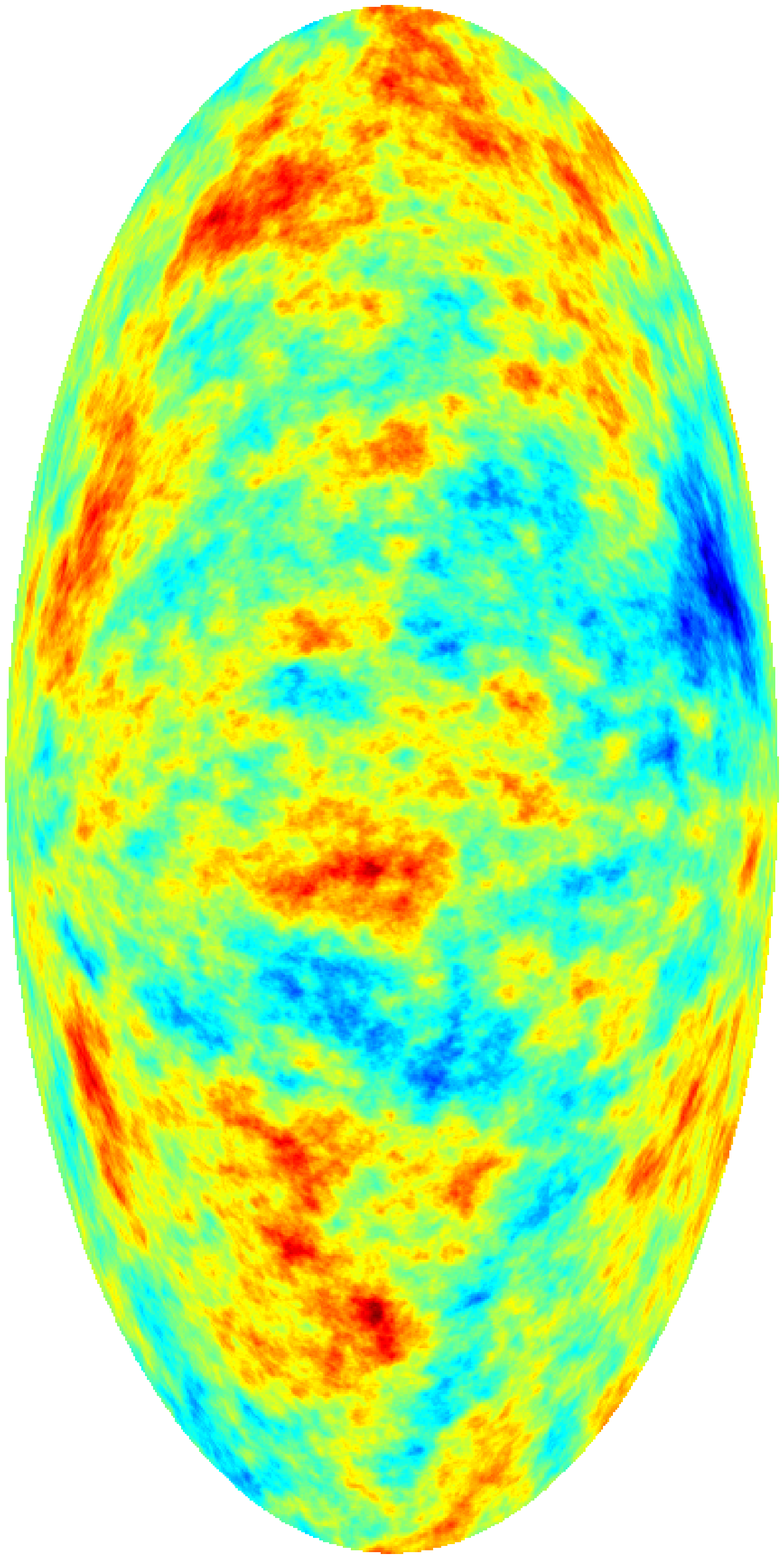}
        \end{turn}
     \end{minipage}
     \begin{minipage}{0.49\linewidth}
        \begin{turn}{90}
       \hspace*{-0cm}\includegraphics[width=0.62\linewidth]{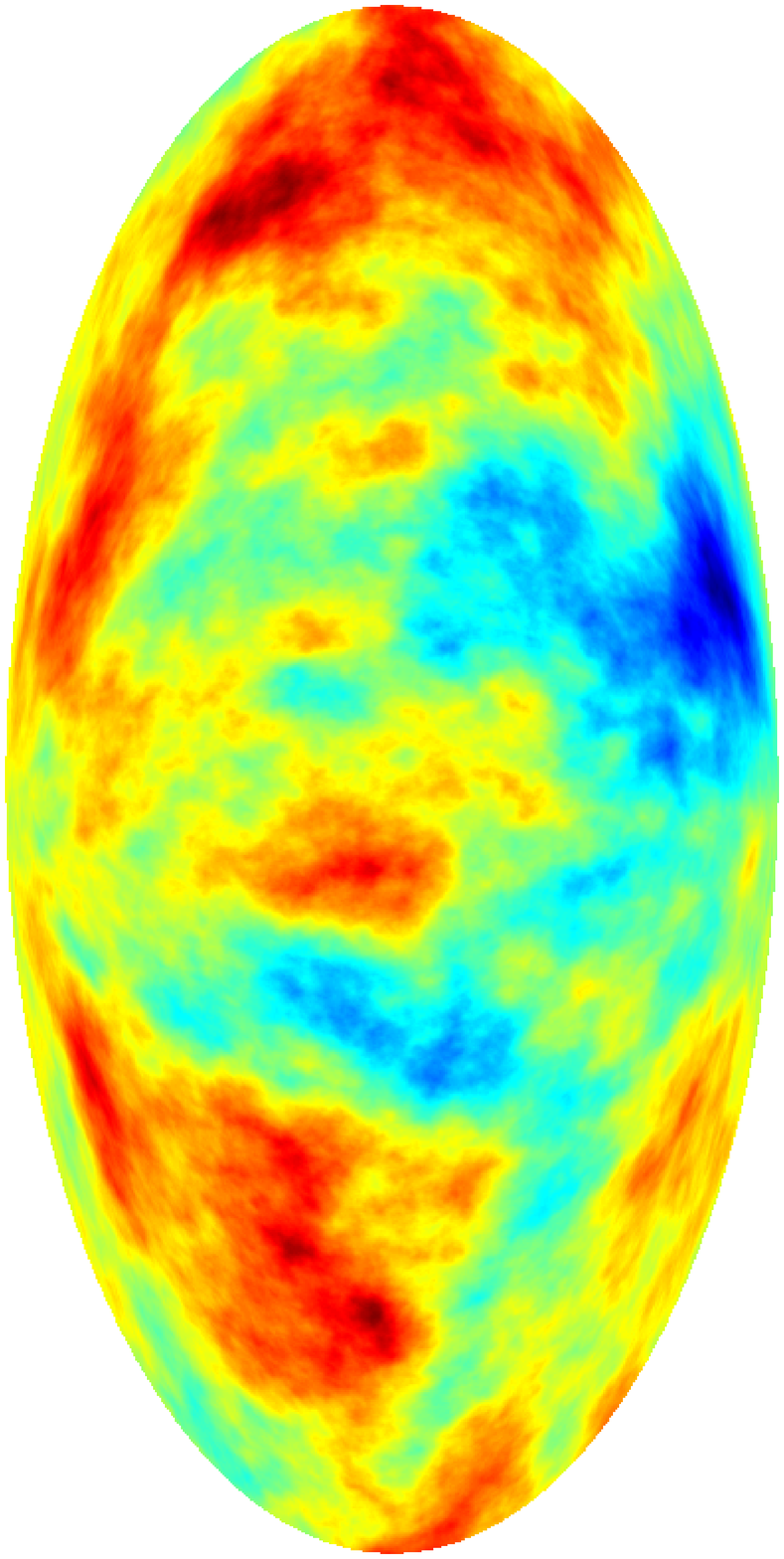}
          \end{turn}
     \end{minipage}

   \end{minipage}

   \caption{Sky maps of the lensing deflection $l(l+1)C_l^\Pi$ with the dipole included, for $m_\nu = 10^{-5}$ eV
   (top-left), $10^{-3}$ eV (top-right), $10^{-2}$ eV (bottom-left) and $10^{-1}$ eV (bottom-right).}
   \label{fig:lensing_sky}
\end{figure}

\begin{figure}
   \noindent
  \begin{minipage}{1.0\linewidth}

     \begin{minipage}{0.55\linewidth}
        \begin{turn}{0}
       \hspace*{-0.4cm}\includegraphics[width=1.0\linewidth]{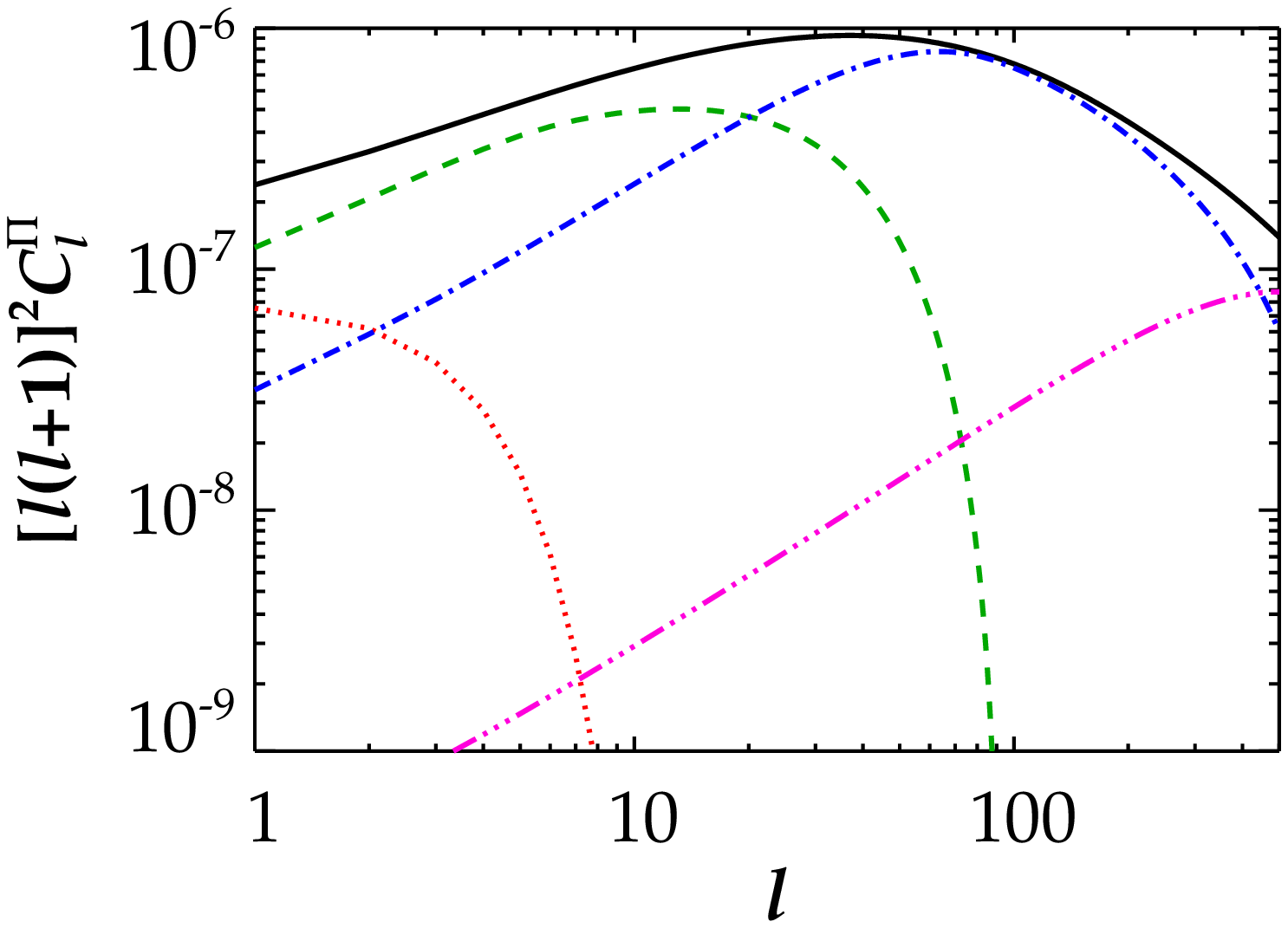}
          \end{turn}
     \end{minipage}
     \begin{minipage}{0.55\linewidth}
        \begin{turn}{0}
       \hspace*{-1.0cm}\includegraphics[width=1.0\linewidth]{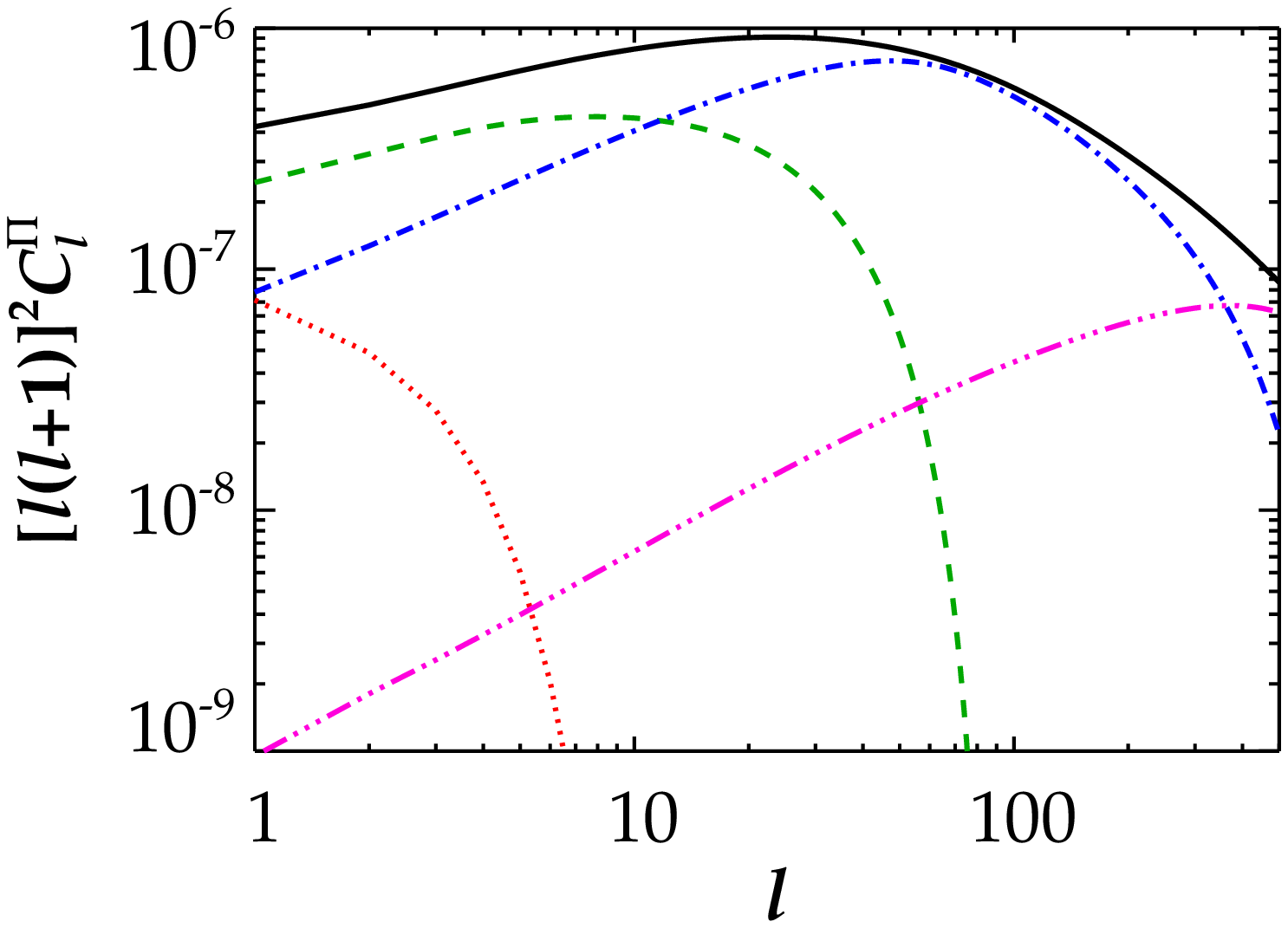}
          \end{turn}
     \end{minipage}
     \begin{minipage}{0.55\linewidth}
        \begin{turn}{0}
        \hspace*{-0.4cm}\includegraphics[width=1.0\linewidth]{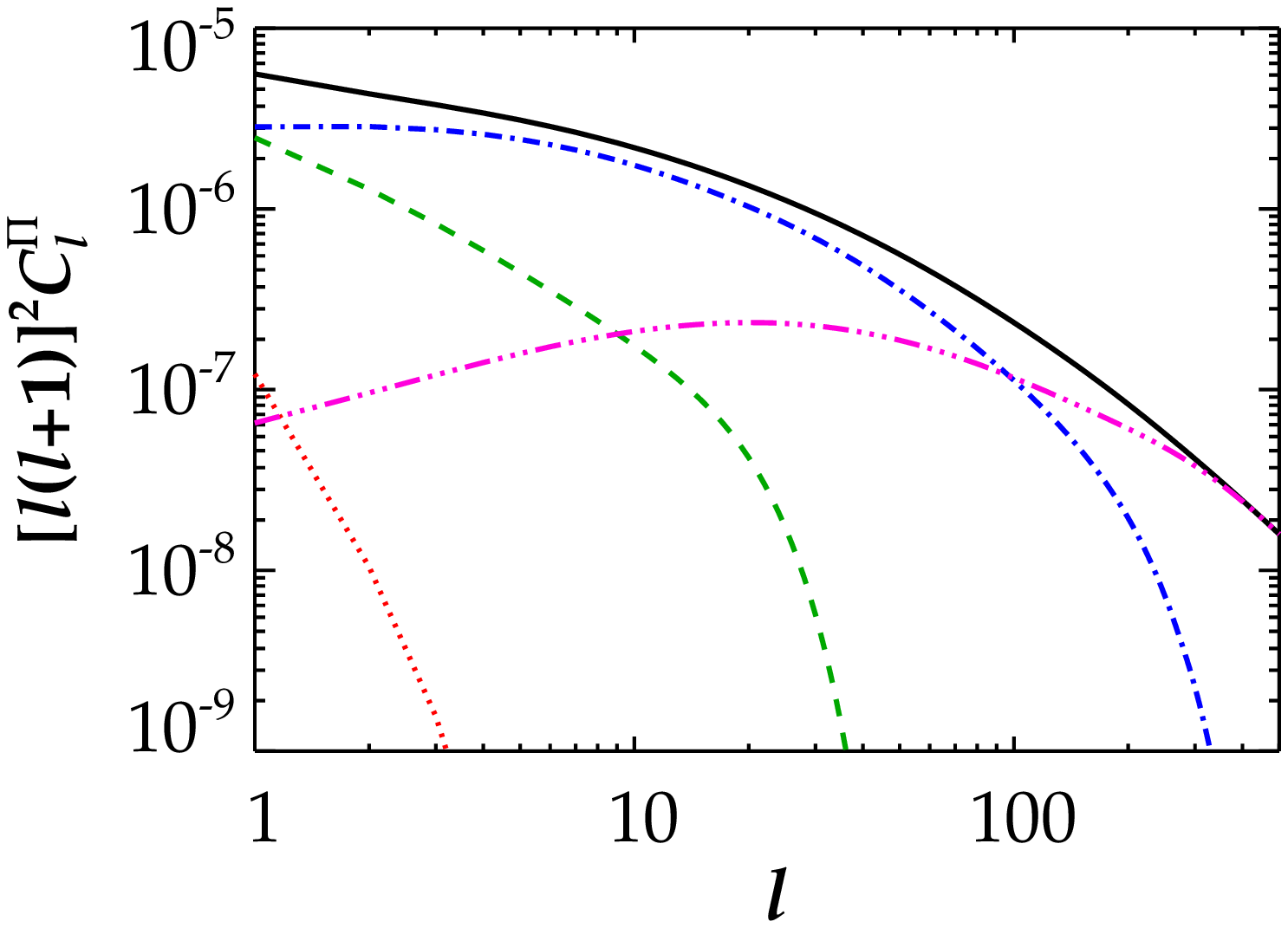}
        \end{turn}
     \end{minipage}
     \begin{minipage}{0.55\linewidth}
        \begin{turn}{0}
       \hspace*{-1.0cm}\includegraphics[width=1.0\linewidth]{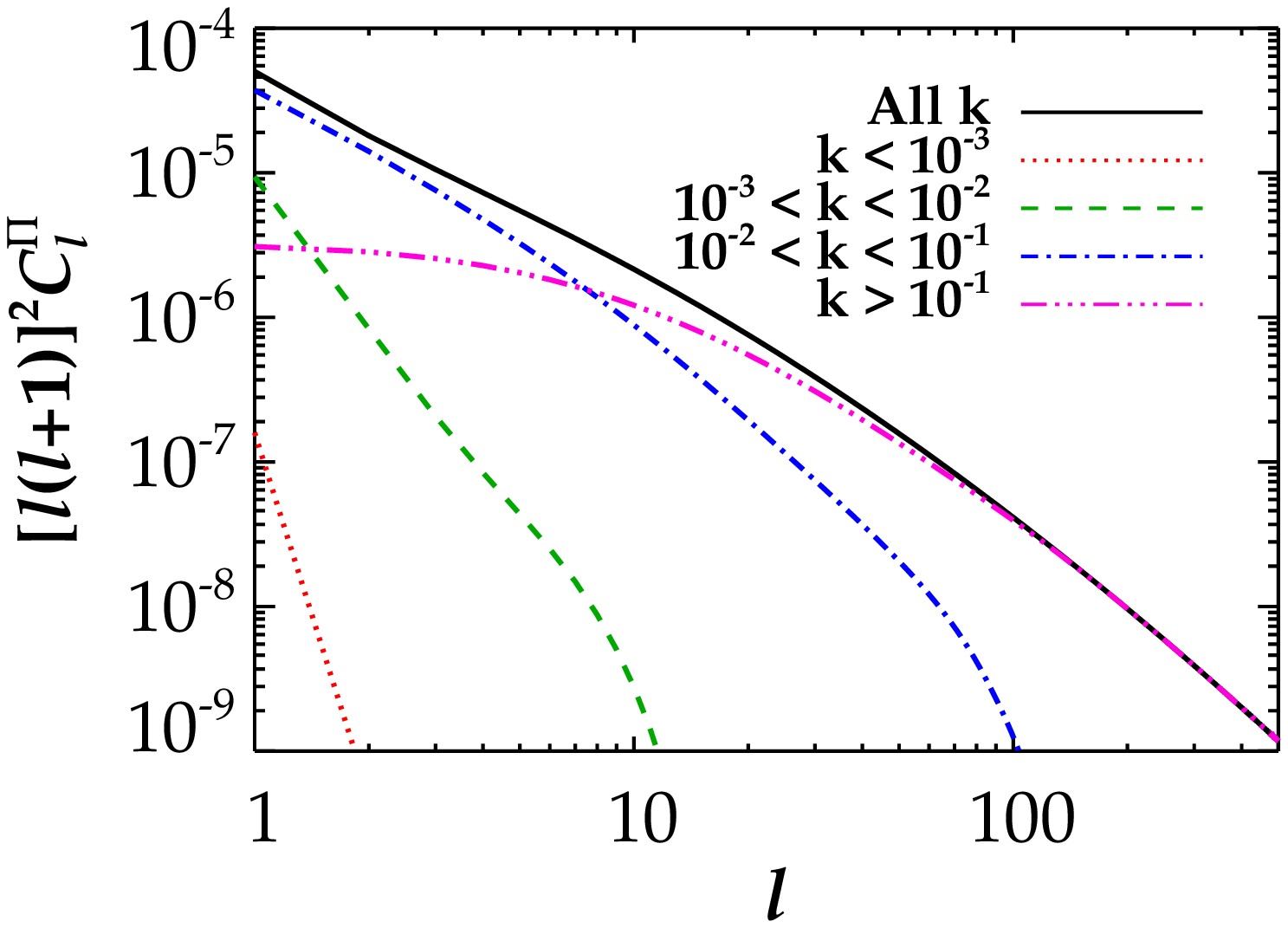}
          \end{turn}
     \end{minipage}
   \end{minipage}

   \caption{The contribution to $C_l^\Pi$ from various scales $k$ in units of $h \, {\rm Mpc}^{-1}$. From
   top-left to bottom-right the neutrino mass is $10^{-5}$ eV, $10^{-3}$ eV, $10^{-2}$ eV and $10^{-1}$ eV.}
   \label{fig:k}
\end{figure}

\subsection{Numerical results}

In Fig.~\ref{fig:lensing} we show $l^2 (l+1)^2 C_l^\Pi$\footnote{Compared to $C_l^\Theta$ their is an extra
factor of $l(l+1)$ since the physically relevant quantity is the deflection angle.} for various neutrino
masses, all in a $\Lambda$CDM background model, and Fig.~\ref{fig:lensing_sky} shows realisations of these spectra. We have done the calculation in linear theory only. As can be
seen, for higher neutrino masses the lensing distortion peaks at lower $l$ because a given $k$-scale
corresponds to lower $l$ when $v < c$. Basically there is no contribution from modes with $k \lesssim
l/\chi^*$. This also means that for higher masses $C_l^\Pi$ picks up a much larger contribution from high $k$. 
This can be seen explicitly in Fig.~\ref{fig:k} which shows the contribution to
$C_l^\Pi$ from various scales. For the higher masses there are significant high-$k$ contributions to lensing
already at low $l$.

From the Limber approximation it is also straightforward to understand how $C_l^\Pi$ changes with neutrino
mass. At high $l$, very approximately we can set $v \propto \chi$ in Eq.~(\ref{eq:limber}), which means that $\frac{\chi}{v}
\left[\frac{\chi^*-\chi}{\chi^* \chi}\right]^2 \propto 1/\chi^2$. The potential power spectrum changes from
$\mathcal{P}_\psi(l/\chi,\eta) \sim$ {\it Const.} at low $l$ to $\mathcal{P}_\psi(l/\chi,\eta) \sim \chi^4$ at high
$l$. Thus, at high $l$ the integrand is proportional to $\chi^2$ which is proportional to $m_\nu^{-2}$. This
explains the lower overall lensing power at high $l$ for high masses.

From Figs.~\ref{fig:transfer} and \ref{fig:lensing} it can be seen that there is a large cross-correlation
between $C_l^\Theta$ and $C_l^\Pi$ at low $l$. This was also shown in \cite{Lewis1}.

\begin{figure}
   \noindent
  \begin{minipage}{1.0\linewidth}
     \begin{minipage}{0.55\linewidth}
        \begin{turn}{0}
        \hspace*{-0.4cm}\includegraphics[width=1\linewidth]{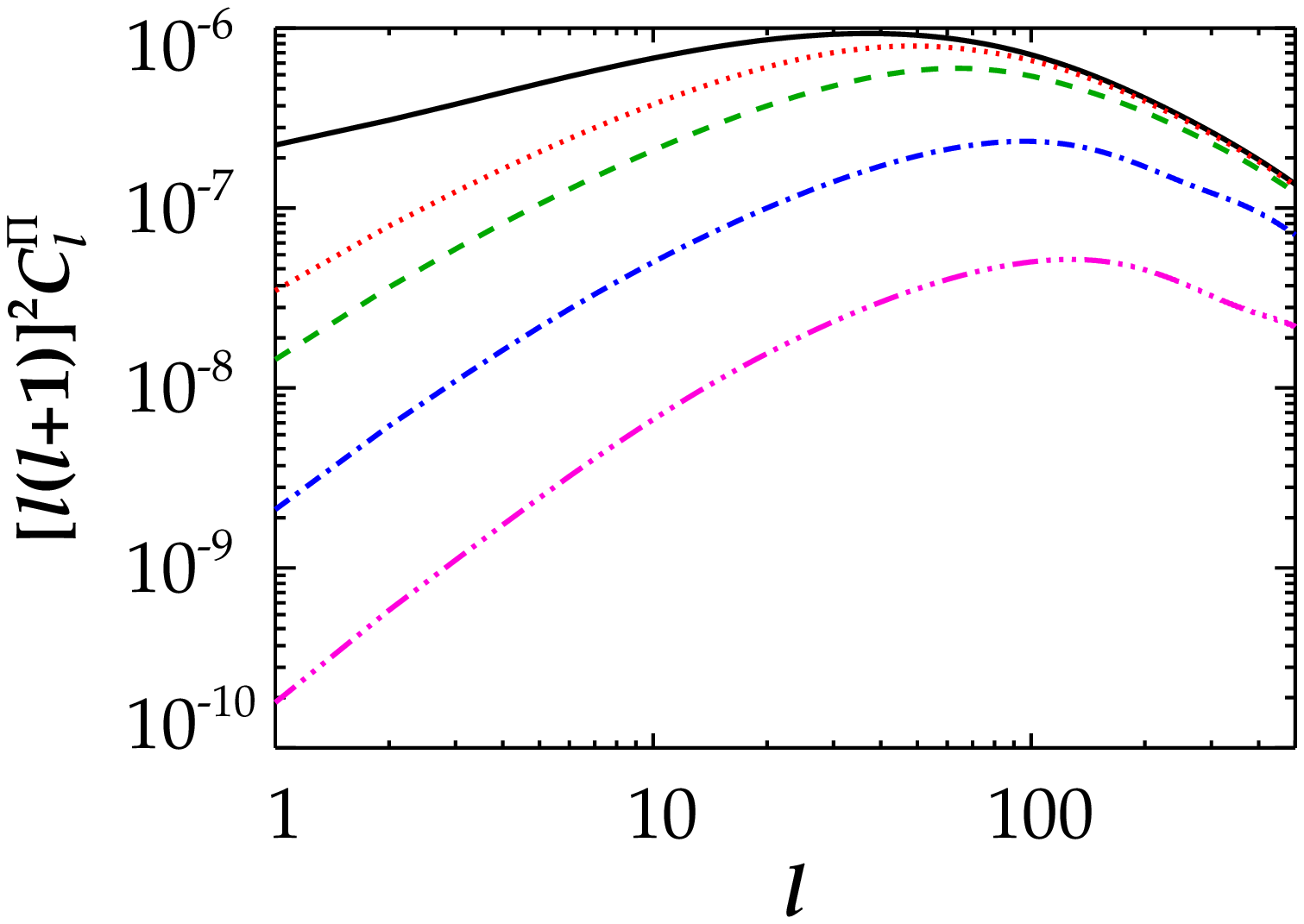}
        \end{turn}
     \end{minipage}
     \begin{minipage}{0.55\linewidth}
        \begin{turn}{0}
        \hspace*{-1.0cm}\includegraphics[width=1\linewidth]{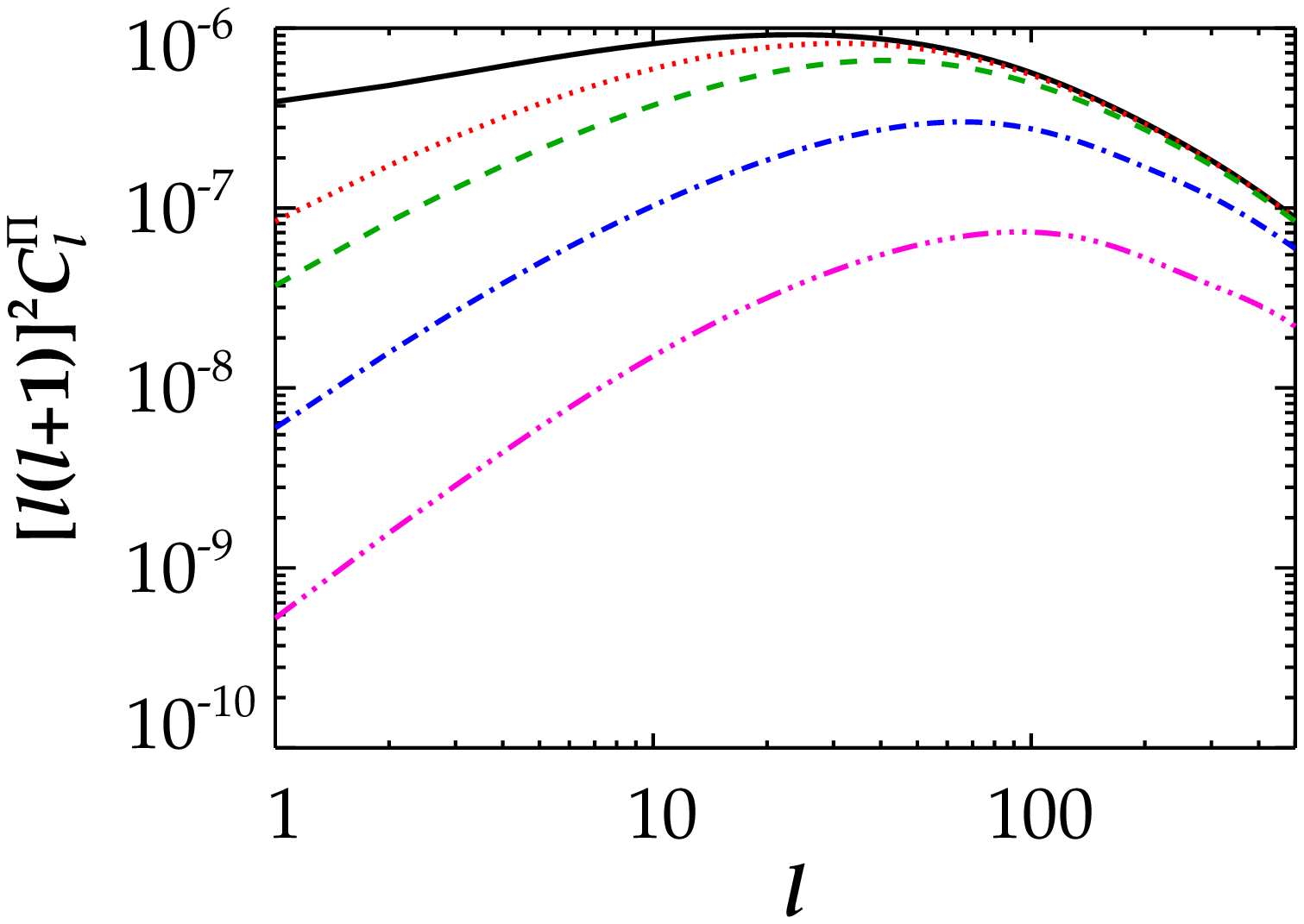}
        \end{turn}
     \end{minipage}
     \begin{minipage}{0.55\linewidth}
        \begin{turn}{0}
       \hspace*{-0.4cm}\includegraphics[width=1\linewidth]{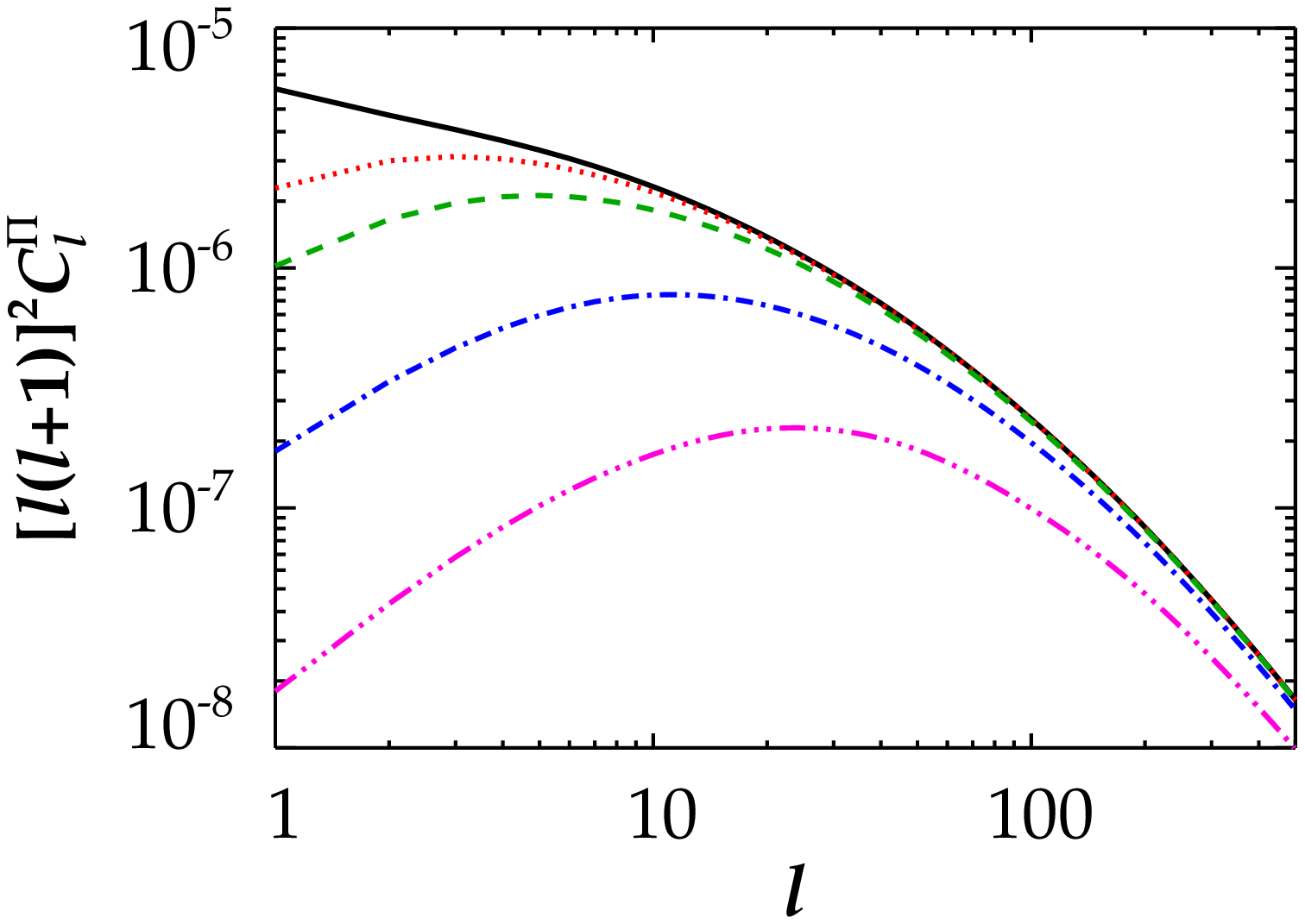}
          \end{turn}
     \end{minipage}
     \begin{minipage}{0.55\linewidth}
        \begin{turn}{0}
        \hspace*{-1.0cm}\includegraphics[width=1\linewidth]{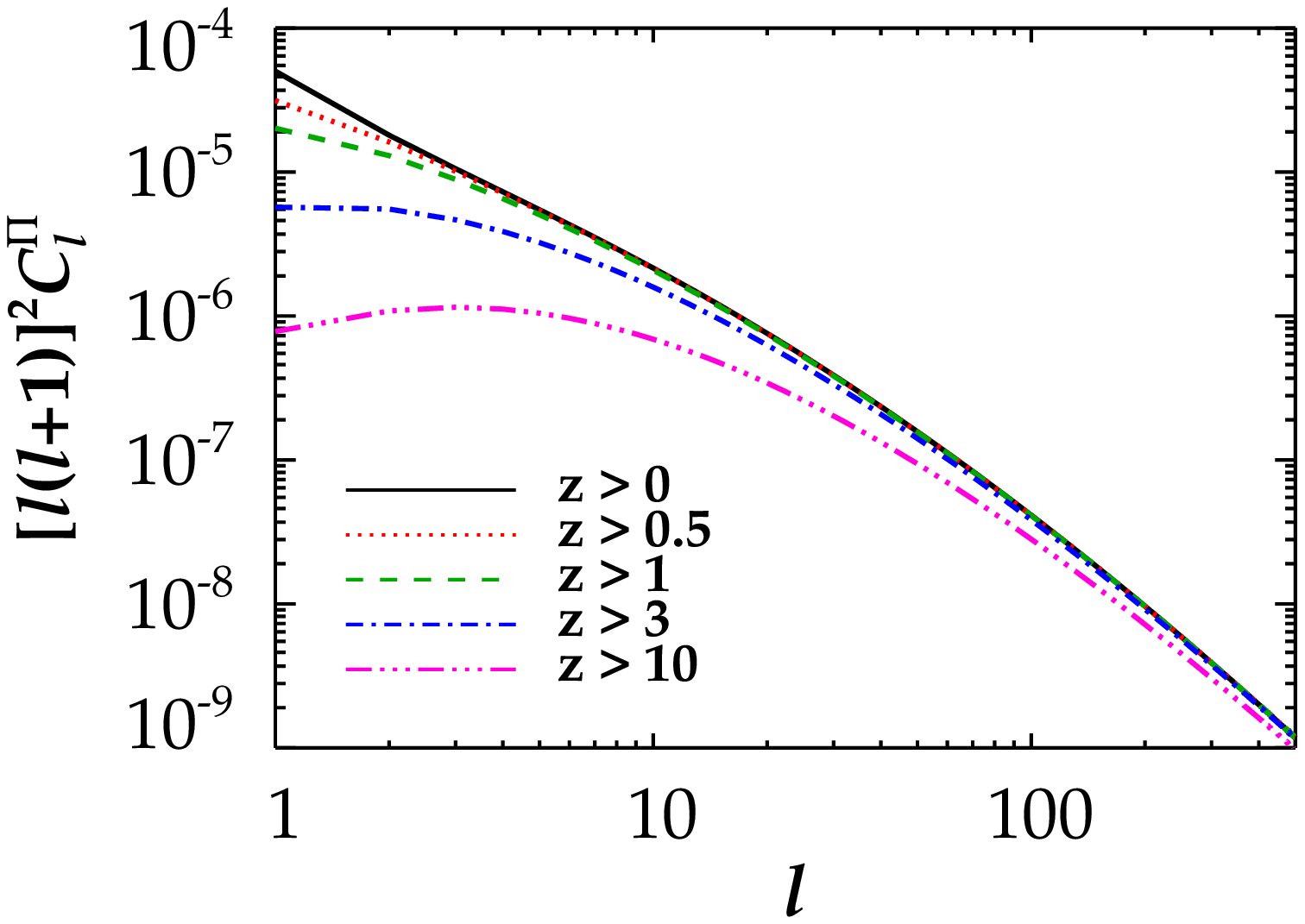}
        \end{turn}
     \end{minipage}
   \end{minipage}

   \caption{Contributions to the lensing potential, $C_l^\Pi$, at different redshifts. From top-left to
   bottom-right the neutrino mass is $10^{-5}$ eV, $10^{-3}$ eV, $10^{-2}$ eV and $10^{-1}$ eV.}
   \label{fig:z}
\end{figure}

In Fig.~\ref{fig:z} we show the contribution to lensing from different redshifts. Almost all of the low $l$
contribution comes at very low $z$. Note that for the $m_\nu = 10^{-2}$ eV case which was semi-relativistic
until a fairly low redshift the high-$z$ contribution is fairly similar to the massless case, i.e.\ the
transition from relativistic to non-relativistic can be seen directly from the change in shape of the lensing
spectrum.

\begin{figure}
   \noindent
      \begin{center}
      \hspace*{-0.1cm}\includegraphics[width=0.8\linewidth]{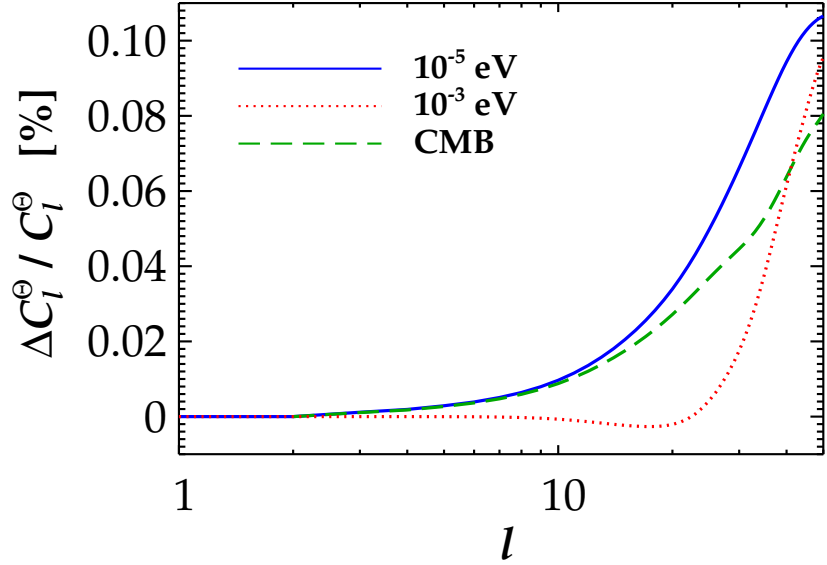}
      \end{center}

   \caption{Percentage difference between the lensed and unlensed angular power spectra,
   $(\tilde{C}_l^\Theta-C_l^\Theta)/C_l^\Theta$, for 2 different neutrino masses and compared with the
   lensing effect on the CMB. The spectra have been slightly smoothed due to finite numerical resolution.}
   \label{fig:diff}
\end{figure}

\section{The lensed C$\nu$B}

\subsection{Theory}

Finally, in this section we combine the results from the two previous sections to derive the lensed {\cnub}
spectrum.

Modelling weak gravitational lensing as a second-order effect, the full-sky lensed angular neutrino power
spectrum is found from
\begin{equation}
\tilde{C}_l^\Theta(q) = 2\pi\int_{-1}^{1}{\tilde{\xi}(q,\beta)d^l_{00}(\beta){\rm d~cos}\beta},
\end{equation}
where $\tilde{\xi}(q,\beta)$ is the lensed correlation function and $\rm{cos}\beta = \hat{\bf{n}}_1 \cdot
\hat{\bf{n}}_2$, where $\hat{\bf{n}}_1$ and $\hat{\bf{n}}_2$ indicate two directions on the sky.
$d^l_{00}(\beta)$ is a special case of the reduced Wigner functions, $d^l_{mm'}(\beta)$, given by
\begin{eqnarray}
d^l_{mm'}(\beta) & = & (-1)^{l-m'}[(l+m)!(l-m)!(l+m')!(l-m')!]^{1/2}\\ \nonumber && \,
\sum_k{(-1)^k\frac{[{\rm cos}(\beta/2)]^{m+m'+2k}[{\rm
sin}(\beta/2)]^{2l-m-m'-2k}}{k!(l-m-k)!(l-m'-k)!(m+m'+k)!}},
\end{eqnarray}
where the sum is over all $k$ fulfilling the criterion that the arguments of the factorials should be
non-negative.

Taking sky curvature into account $\tilde{\xi}(q,\beta)$ is given by
\begin{eqnarray}
\tilde{\xi}(q,\beta) & \simeq & \sum_{l=1}^\infty\frac{2l+1}{4\pi}C_l^\Theta(q) e^{-L^2C_+(q,0)/2}\\
\nonumber && \,
\sum_{mm'}{d^l_{mm'}(\beta)I_{\frac{m+m'}{2}}[L^2C_+(q,\beta)/2]I_{\frac{m-m'}{2}}[L^2C_-(q,\beta)/2]}.
\label{eq:tilde_xi}
\end{eqnarray}
Here $C_l^\Theta(q)$ is the unlensed power spectrum and in the double sum, $\sum_{mm'}$, $m$ and $m'$ runs
from $-~l$ to $+~l$ in integer steps with the criterion that $m+m'$ is even. $L = l + 1/2$ and $I_n(x)$ is the modified
Bessel function of the first kind. Finally
\begin{equation}
C_\pm(q,\beta) = \sum_{l=1}^\infty{\frac{2l+1}{4\pi}l(l+1)C_l^\Pi(q) d_{\pm11}^l(\beta)},
\end{equation}
where $C_l^\Pi(q)$ is the power spectrum of the lensing potential.

\subsection{Numerical results}

We have found $\tilde{C}_l^\Theta$ with the lensing code in CAMB and Fig.~\ref{fig:diff}
shows the difference with respect to the unlensed spectra for the $m_\nu = 10^{-5}$ eV and $10^{-3}$ eV
cases, as well as for the CMB. Both neutrino spectra are closely correlated with the CMB case because the
main effect is to move power from the higher $l$ modes in the primary spectra which are very similar at
these $l$-values. Since the primary spectrum dips to a minimum at a somewhat higher $l$ for $m_\nu=10^{-3}$ eV
than for $m_\nu=10^{-5}$ eV the lensing effect also kicks in later, as can be seen in Fig.~\ref{fig:diff}.

We have not presented lensing results for the higher masses because the relative effect of lensing becomes
less important at low $l$, and because numerical noise from the truncation of both primary and lensing spectra
at $l=500$ prevents a reliable calculation of the lensed {\cnub} spectrum much beyond $l \sim 10$ for the most massive cases.

Note that we have found $\tilde{C}_l^\Theta$ by lensing the average primary spectrum with the average lensing
spectrum. For small neutrino masses with $v \simeq c$ this is surely a good approximation. For higher masses
it will be more important to lense each $C_l^\Theta(q)$ with its corresponding $C_l^\Pi(q)$ and then calculate
an energy average, though the order of averaging should not significantly affect the total $\tilde{C}_l^\Theta$.

\section{Discussion and conclusions}

We have calculated the anisotropy of the {\cnub} in linear theory which applies to neutrino masses of less than
$\sim 0.1$ eV. For massless neutrinos the power spectrum of {\cnub} fluctuations closely resembles the usual CMB
spectrum, but with the baryon-photon acoustic oscillations absent.

At high $l$ the neutrino spectra are almost identical, independent of the neutrino mass. The reason is that at
high $l$ all neutrinos are dominated by free-streaming which in $l$-space has approximately the same impact for all masses.

For smaller $l$-values the anisotropy increases dramatically as the mass increases, because the gravitational
source term becomes much more important at late times for massive particles. This initially increases the lowest
multipoles but via the Boltzmann hierarchy the effect quickly propagates to higher $l$.

We then proceeded to calculate the effect of weak gravitational lensing for massive neutrinos and found it to
be much stronger at low $l$, and correspondingly weaker at high $l$, as compared to the massless case. Finally we calculated the effect of lensing
on the primary {\cnub} spectra and found the effect to be unimportant (with relative changes at the per mille
level up to $l \sim 50$), but with some differences depending on the neutrino mass.

 It is worth mentioning that any direct experimental measurement of the {\cnub} anisotropy will most likely measure flavour states, not mass states. The actual anisotropy measured will therefore be a superposition of anisotropies for three different mass states, weighed with their individual flavour content.

We should finally again stress that our results are only valid for masses of $\lwig 0.1$ eV. For higher masses
linear perturbation
theory breaks down because neutrino streaming velocities become comparable to the typical gravitational flow
velocities so that a significant fraction of neutrinos are bound in structures. In this case the {\cnub}
spectrum must be found from $N$-body simulations of neutrino clustering \cite{Brandbyge1}. This can also be seen from the fact
that the anisotropy at low $l$ is a factor $\sim 10^9$ higher for $m_\nu=0.1$ eV than for massless neutrinos.
Since the anisotropy for massless particles corresponds to $\delta \rho/\rho \sim 10^{-5}$ the corresponding
$\delta \rho/\rho$ for 0.1 eV neutrinos is of order one, indicating that perturbation theory breaks down.

\section*{Acknowledgements}
We acknowledge computing resources from the Danish Center for Scientific Computing (DCSC).

\section*{References} 



\begin{thebibliography}{00}

\bibitem{Komatsu:2008hk}
  E.~Komatsu {\it et al.}  [WMAP Collaboration],
  Astrophys.\ J.\ Suppl.\  {\bf 180}, 330 (2009)
  [arXiv:0803.0547 [astro-ph]].

\bibitem{Hamann:2007pi}
  J.~Hamann, S.~Hannestad, G.~G.~Raffelt and Y.~Y.~Y.~Wong,
  JCAP {\bf 0708}, 021 (2007)
  [arXiv:0705.0440 [astro-ph]].

\bibitem{de Bernardis:2007bu}
  F.~de Bernardis, A.~Melchiorri, L.~Verde and R.~Jimenez,
  JCAP {\bf 0803}, 020 (2008)
  [arXiv:0707.4170 [astro-ph]].


\bibitem{Ichikawa:2008pz}
  K.~Ichikawa, T.~Sekiguchi and T.~Takahashi,
  Phys.\ Rev.\  D {\bf 78}, 083526 (2008)
  [arXiv:0803.0889 [astro-ph]].

\bibitem{Hamann:2008we}
  J.~Hamann, S.~Hannestad, A.~Melchiorri and Y.~Y.~Y.~Wong,
  JCAP {\bf 0807}, 017 (2008)
  [arXiv:0804.1789 [astro-ph]].

\bibitem{Popa:2008tb}
  L.~A.~Popa and A.~Vasile,
  JCAP {\bf 0806}, 028 (2008)
  [arXiv:0804.2971 [astro-ph]].


\bibitem{Bashinsky:2003tk}
  S.~Bashinsky and U.~Seljak,
  Phys.\ Rev.\  D {\bf 69}, 083002 (2004)
  [arXiv:astro-ph/0310198].

\bibitem{Trotta:2004ty}
  R.~Trotta and A.~Melchiorri,
  Phys.\ Rev.\ Lett.\  {\bf 95}, 011305 (2005)
  [arXiv:astro-ph/0412066].

\bibitem{Bell:2005dr}
  N.~F.~Bell, E.~Pierpaoli and K.~Sigurdson,
  Phys.\ Rev.\  D {\bf 73}, 063523 (2006)
  [arXiv:astro-ph/0511410].

\bibitem{DeBernardis:2008ys}
  F.~De Bernardis, L.~Pagano, P.~Serra, A.~Melchiorri and A.~Cooray,
  JCAP {\bf 0806}, 013 (2008)
  [arXiv:0804.1925 [astro-ph]].

\bibitem{Basboll:2008fx}
  A.~Basboll, O.~E.~Bjaelde, S.~Hannestad and G.~G.~Raffelt,
  Phys.\ Rev.\  D {\bf 79}, 043512 (2009)
  [arXiv:0806.1735 [astro-ph]].

\bibitem{Hannestad:2004qu}
  S.~Hannestad,
  JCAP {\bf 0502}, 011 (2005)
  [arXiv:astro-ph/0411475].

\bibitem{Friedland:2007vv}
  A.~Friedland, K.~M.~Zurek and S.~Bashinsky,
  arXiv:0704.3271 [astro-ph].

 \bibitem{Weinberg:1962zz}
  S.~Weinberg,
  Phys.\ Rev.\  {\bf 128}, 1457 (1962).

\bibitem{Cocco:2007za}
  A.~G.~Cocco, G.~Mangano and M.~Messina,
  JCAP {\bf 0706}, 015 (2007)
  [J.\ Phys.\ Conf.\ Ser.\  {\bf 110}, 082014 (2008)]
  [arXiv:hep-ph/0703075].

\bibitem{Blennow:2008fh}
  M.~Blennow,
  Phys.\ Rev.\  D {\bf 77}, 113014 (2008)
  [arXiv:0803.3762 [astro-ph]].


 \bibitem{Weiler:1982qy}
  T.~J.~Weiler,
  Phys.\ Rev.\ Lett.\  {\bf 49}, 234 (1982).

 \bibitem{Stodolsky:1974aq}
  L.~Stodolsky,
  Phys.\ Rev.\ Lett.\  {\bf 34}, 110 (1975)
  [Erratum-ibid.\  {\bf 34}, 508 (1975)].

 \bibitem{Gelmini:2004hg}
  G.~B.~Gelmini,
  Phys.\ Scripta {\bf T121}, 131 (2005)
  [arXiv:hep-ph/0412305].

\bibitem{Ringwald:2004np}
  A.~Ringwald and Y.~Y.~Y.~Wong,
  JCAP {\bf 0412}, 005 (2004)
  [arXiv:hep-ph/0408241].

 \bibitem{Fodor:2002hy}
  Z.~Fodor, S.~D.~Katz and A.~Ringwald,
  JHEP {\bf 0206}, 046 (2002)
  [arXiv:hep-ph/0203198].

 \bibitem{Duda:2001hd}
  G.~Duda, G.~Gelmini and S.~Nussinov,
  Phys.\ Rev.\  D {\bf 64}, 122001 (2001)
  [arXiv:hep-ph/0107027].

 \bibitem{Langacker:1982ih}
  P.~Langacker, J.~P.~Leveille and J.~Sheiman,
  Phys.\ Rev.\  D {\bf 27}, 1228 (1983).

\bibitem{Cabibbo:1982bb}
  N.~Cabibbo and L.~Maiani,
  Phys.\ Lett.\  B {\bf 114}, 115 (1982).


\bibitem{Pastor:2001iu}
  S.~Pastor, G.~G.~Raffelt and D.~V.~Semikoz,
  Phys.\ Rev.\  D {\bf 65}, 053011 (2002)
  [arXiv:hep-ph/0109035].

\bibitem{Pastor:2008ti}
  S.~Pastor, T.~Pinto and G.~G.~Raffelt,
  Phys.\ Rev.\ Lett.\  {\bf 102}, 241302 (2009)
  [arXiv:0808.3137 [astro-ph]].

\bibitem{Simha:2008mt}
  V.~Simha and G.~Steigman,
  JCAP {\bf 0808}, 011 (2008)
  [arXiv:0806.0179 [hep-ph]].

\bibitem{Wong:2002fa}
  Y.~Y.~Y.~Wong,
  Phys.\ Rev.\  D {\bf 66}, 025015 (2002)
  [arXiv:hep-ph/0203180].

\bibitem{Abazajian:2002qx}
  K.~N.~Abazajian, J.~F.~Beacom and N.~F.~Bell,
  Phys.\ Rev.\  D {\bf 66}, 013008 (2002)
  [arXiv:astro-ph/0203442].



\bibitem{Hu:1995fqa}
  W.~Hu, D.~Scott, N.~Sugiyama and M.~J.~.~White,
  Phys.\ Rev.\  D {\bf 52}, 5498 (1995)
  [arXiv:astro-ph/9505043].

\bibitem{Michney:2006mk}
  R.~J.~Michney and R.~R.~Caldwell,
  JCAP {\bf 0701}, 014 (2007)
  [arXiv:astro-ph/0608303].

\bibitem{MB}
  C.~P.~Ma and E.~Bertschinger,
  Astrophys.\ J.\  {\bf 455}, 7 (1995)
  [arXiv:astro-ph/9506072].

 \bibitem{CAMB}
  A.~Lewis and S.~Bridle,
  Phys.\ Rev.\ D {\bf 66}, 103511 (2002)
  [arXiv:astro-ph/0205436].

\bibitem{Bertschinger:1995er}
  E.~Bertschinger,
  arXiv:astro-ph/9506070.

\bibitem{Gorski:2004by}
  K.~M.~Gorski, E.~Hivon, A.~J.~Banday, B.~D.~Wandelt, F.~K.~Hansen, M.~Reinecke and M.~Bartelman,
  Astrophys.\ J.\  {\bf 622}, 759 (2005)
  [arXiv:astro-ph/0409513].

\bibitem{Dodelson:2009ze}
  S.~Dodelson and M.~Vesterinen,
  arXiv:0907.2887 [astro-ph.CO].

\bibitem{Brandbyge1}
  J.~Brandbyge and S.~Hannestad,
  [arXiv:0908.1969 [astro-ph]].

\bibitem{Seljak:1995ve}
  U.~Seljak,
  Astrophys.\ J.\  {\bf 463}, 1 (1996)
  [arXiv:astro-ph/9505109].

\bibitem{Challinor1}
  A.~Challinor and A.~Lewis,
  Phys.\ Rev.\  D {\bf 71} (2005) 103010
  [arXiv:astro-ph/0502425].

\bibitem{Lewis1}
  A.~Lewis and A.~Challinor,
  Phys.\ Rept.\  {\bf 429} (2006) 1
  [arXiv:astro-ph/0601594].

\bibitem{kaiser}
  N. Kaiser,
  Astrophys.\ J.\  {\bf 388}, 272 (1992)
  [arXiv:astro-ph/9603033].

 \bibitem{Jain:1996st}
  B.~Jain and U.~Seljak,
  Astrophys.\ J.\  {\bf 484}, 560 (1997)
  [arXiv:astro-ph/9611077].

\end{thebibliography}
\end{document}